\documentclass[preprint,authoryear,11pt]{elsarticle}

\usepackage{graphics}
\usepackage{graphicx}
\usepackage{epsfig}
\usepackage{subfigure}
\usepackage{amssymb}
\usepackage{amsthm}
\usepackage{bm}
\usepackage{amsmath}
\usepackage{color}
\usepackage{geometry}
\usepackage{setspace}

\allowdisplaybreaks
\usepackage{enumerate}
\usepackage{booktabs}

\usepackage{hyperref}
\hypersetup{
	colorlinks=true,       
}

\biboptions{}

\journal{}

\begin{document}

\begin{frontmatter}



\title{Nonlinear response and axisymmetric wave propagation \\ in functionally graded soft electro-active tubes}


\author[1]{Bin Wu}
\author[1,2]{Michel Destrade}
\author[2,3,4]{Weiqiu Chen\corref{cor1}}
\ead{chenwq@zju.edu.cn}

\cortext[cor1]{Corresponding author. Tel.: +86-571-87951866; fax: +86-571-87951866.}
\address[1]{School of Mathematics, Statistics and Applied Mathematics,\\ NUI Galway, University Road, Galway, Ireland; \\[6pt]}

\address[2]{Key Laboratory of Soft Machines and Smart Devices of Zhejiang Province \\ and Department of Engineering Mechanics, \\ Zhejiang University, Hangzhou 310027, P.R. China;\\[6pt]}

\address[3]{State Key Lab of CAD \& CG,
Zhejiang University, Hangzhou 310058, P.R. China;\\[6pt]}

\address[4]{Soft Matter Research Center, Zhejiang University, Hangzhou 310027, P.R. China.}

\begin{abstract}
	
Soft electro-active (SEA) materials can be designed and manufactured with gradients in their material properties, to modify and potentially improve their mechanical response in service. 
Here, we investigate the nonlinear response of, and axisymmetric wave propagation in a soft circular tube made of a functionally graded SEA material and subject to several biasing fields, including axial pre-stretch, internal/external pressure, and through-thickness electric voltage. 
We take the energy density function of the material to be of the Mooney-Rivlin ideal dielectric type, with material parameters changing linearly along the radial direction. 
We employ the general theory of nonlinear electro-elasticity  to obtain explicitly the nonlinear response of the tube to the applied fields. 
To study wave propagation under inhomogeneous biasing fields, we formulate the incremental equations of motion within the state-space formalism. 
We adopt the approximate laminate technique to derive the analytical dispersion relations for the small-amplitude torsional and longitudinal waves superimposed on a finitely deformed state. Comprehensive numerical results then illustrate that the material gradients and biasing fields have significant influences on the static nonlinear response and on the axisymmetric wave propagation in the tube. 
This study lays the groundwork for designing SEA actuators with improved performance, for tailoring tunable SEA waveguides, and for characterizing non-destructively  functionally graded tubular structures.

\end{abstract}


\end{frontmatter}




\section{Introduction}


As a new type of inhomogeneous materials, functionally graded materials (FGMs) are characterized by  gradual variations in composition and structure over their volume, resulting in continuous changes in their mechanical, physical, or biological properties along one or more directions. Compared with  traditional materials, FGMs may display some unprecedented performance such as reducing residual stress and thermal stress, enhancing fracture toughness of interface bonding, eliminating sharp stress discontinuities, and improving efficiency and life of acoustic wave devices \citep{birman2007modeling, zhong2012mechanics}. 

Historically, the concept of FGMs was first proposed in the mid-1980s by a group of Japanese material scientists, who manufactured a functionally graded thermal barrier capable of withstanding a large temperature gradient. 
Since then, FGMs have been widely used not only in the field of aerospace engineering but also in biology \citep{miyamoto2013functionally, pompe2003functionally}. 
No less than fourteen different fabrication approaches of FGMs are described in detail in the book by \citet{schwartzemerging}, including bulk particulate processing, preform processing, layer processing, and melt processing. 
In the context of linear elasticity, a considerable amount of literature has focused on the theoretical modelling, numerical methods, and experimental testing of FGMs, and the interested reader is referred to the review article by \citet{birman2007modeling} and the book edited by \citet{zhong2012mechanics}.

Although the concept of FGMs is now well established, it is still relatively rare to use it in \textit{soft materials} with highly elastic behavior such as  rubber-like materials and gels. 
A unique superiority of a soft FGM is that it can be exploited to create a smooth transition between a soft and a hard material, thereby eliminating the structural stiffness mismatch. 

Recently, several  soft hyperelastic structures with functionally graded properties have been proposed and prepared.
Hence, making use of a construction-based layering method, \citet{ikeda2002preparation} achieved functionally graded styrene-butadiene rubber vulcanization. 
\citet{libanori2012stretchable} used a hierarchical reinforcement approach to successfully produce a heterogeneous composite with extreme soft-to-hard transition and tunable local elastic stiffness. 
\citet{bartlett20153d} 3D-printed a combustion-powered soft robot, whose explosive actuator connected to pneumatic pegs possesses a stiffness gradient spanning three orders of magnitude. 
\citet{li2019mechanical} used repeated freeze/thaw cycles to create a vertical gradient of mechanical properties in blocs of soft PVA cryogels. 

With respect to a theoretical analysis of soft FGMs, several works have been devoted to the static nonlinear response and elastic wave behaviors of incompressible or compressible functionally graded hyperelastic structures \citep{saravanan2005inflation, batra2009inflation, chen2017bifurcation, wu2018propagation, chen2019tunable, li2019mechanical}. 
Already, the existing research recognizes the critical role played by material gradients in the mechanical response and the function of soft materials and structures.


As one type of promising intelligent materials, \textit{soft electro-active (SEA) materials} have received intensive academic and industrial interest owing to their superior electro-mechanical coupling properties. This is reflected by the following three breakthroughs: (i) A general theoretical framework of nonlinear electro-elasticity has been well developed to describe their high nonlinearity and notable electro-mechanical coupling \citep{mcmeeking2005electrostatic, dorfmann2006nonlinear, dorfmann2014nonlinear, suo2008nonlinear}; (ii) Their advantages, such as rapid response and large deformation under electric stimuli as well as high energy density, have been confirmed experimentally, making SEA materials ideal candidates for broad applications as transducers, actuators, sensors, energy harvesters, biomedical devices and {\color{black}flexible electronics \citep{carpi2011dielectric, anderson2012multi, zhao2014harnessing, lu2020mechanics};} (iii) The application of biasing fields (for example, pre-stretch, internal/external pressure, electric stimuli, etc.) results in significant changes in vibration and wave characteristics of SEA materials, prefiguring various potential prospects in tunable resonators, loudspeakers, vibration isolators, waveguides, phononic crystals and metamaterials \citep{zhu2010resonant, shmuel2012rayleigh, zhao2016application, galich2017shear, wu2018tuning, mao2019electrostatically, chen2020effects}.

Since \citet{pelrine1998electrostriction} proposed an SEA tube actuator, there has been an increasing interest in studying the nonlinear static and dynamic responses of this particular structural configuration subject to biasing fields \citep{dorfmann2006nonlinear, zhu2010large, son2010dynamic, melnikov2016finite, bortot2018nonlinear}. 
Based on the theory of nonlinear electro-elasticity and its linearized incremental theory developed by \citet{dorfmann2006nonlinear, dorfmann2010electroelastic}, considerable efforts have been devoted to elucidating the effects of biasing fields on the structural stability of SEA tubes \citep{zhu2010large, an2015experimental, su2018wrinkles, su2020pattern}, elastic wave propagation \citep{chen2012waves, shmuel2013axisymmetric, shmuel2015manipulating, su2016propagation, wu2017guided}, and tunable {\color{black}vibrations and oscillations \citep{son2010dynamic, sarban2011tubular, bortot2018nonlinear, dorfmann2020waves, Zhu2020}.} 
Note that recent advances in the study of instabilities, tunable phononic crystals, and acoustic/vibration control based on SEA materials have been reviewed in the papers by \citet{zhao2016application}, \citet{dorfmann2019instabilities} and \citet{wang2020tunable}.

To model electrode protection from aggressive agents, achieve greater actuation and improve the efficiency of energy harvesting, some researchers have proposed laminated or stacked SEA actuators and energy harvesters \citep{kovacs2009stacked, tutcuoglu2014energy, calabrese2018active}, which have been  recently studied theoretically by  \citet{gei2013performance}, \citet{cohen2017stacked}, or \citet{bortot2019analysis}.

\begin{figure}[htbp]
	\centering	
	\includegraphics[width=0.75\textwidth]{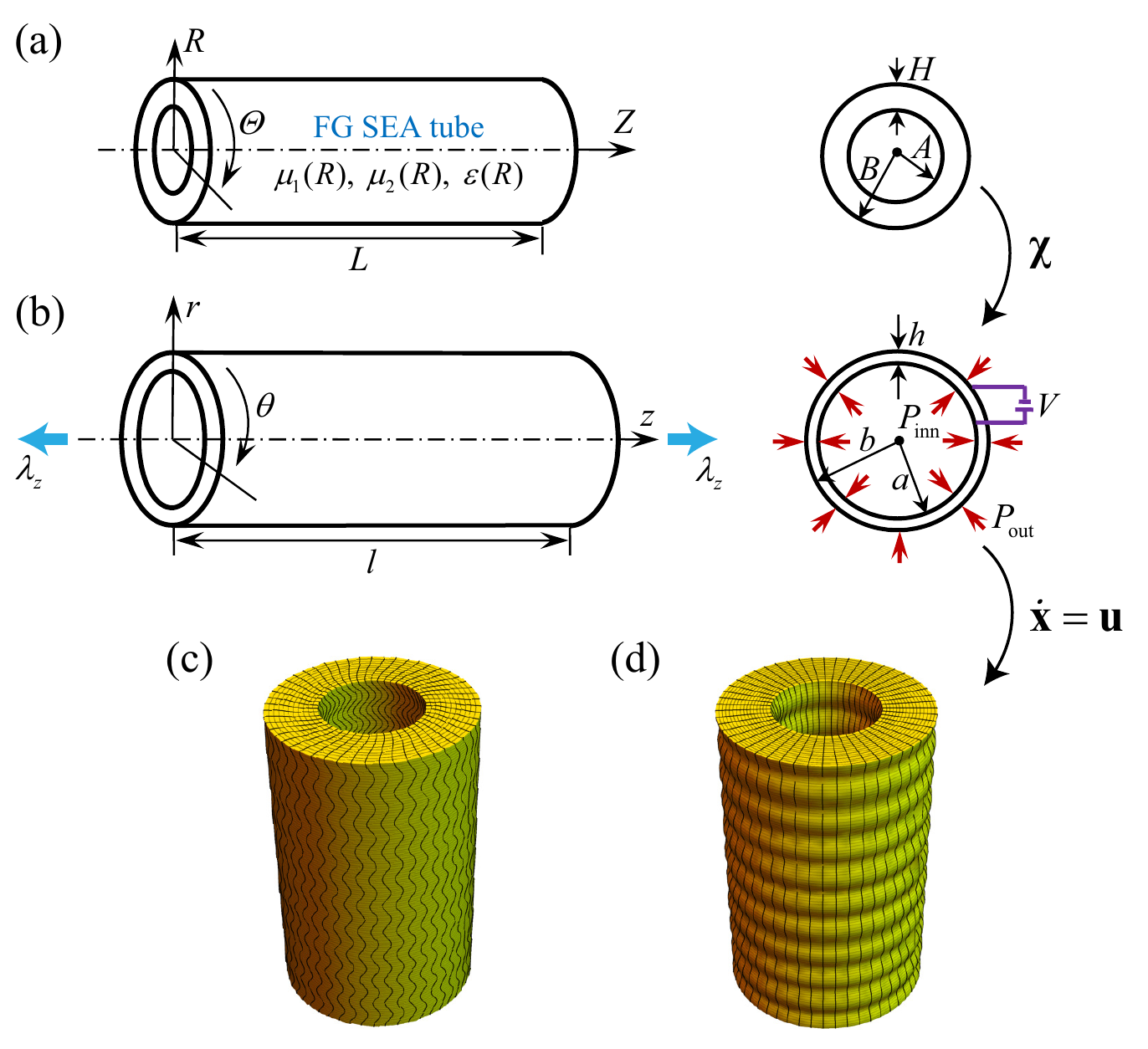}
	\caption{Schematic diagram of a functionally graded SEA tube with its cylindrical coordinates and cross sections: (a) undeformed configuration before activation; (b) deformed configuration after activation induced by a combined action of axial pre-stretch $\lambda_z$, radial electric voltage $V$ as well as internal ($P_{\text{inn}}$) and external ($P_{\text{out}}$) pressures.
	Incremental motion fields created by (c) torsional T-waves and (d) longitudinal L-waves.}
	\label{Fig1}
\end{figure}

The present work introduces the concept of functional gradients of properties into the study of SEA materials, to investigate the nonlinear response of, and elastic guided wave propagation in tubes under coupled biasing fields. 
The tubes are characterized by the Mooney-Rivlin ideal dielectric model with affine variations of material parameters in the radial direction \citep{batra2009inflation, chen2017bifurcation}. The biasing fields are induced by the combined action of an axial pre-stretch, a pressure difference and an electric voltage applied to the electrodes on the inner and outer cylindrical surfaces, resulting in nonlinear axisymmetric deformations as shown in Figs.~\ref{Fig1}(a)-(b). We consider both incremental torsional and longitudinal guided waves (hereafter abbreviated as the T waves and L waves, respectively) along the axial direction in the deformed functionally graded SEA tube (see Figs.~\ref{Fig1}(c)-(d)). To deal with the inhomogeneous biasing fields, we use the state-space method (SSM) \citep{wu2017guided, wu2018propagation} to analyze  axisymmetric elastic wave propagation.

This paper is organized as follows. Section~\ref{section2} summarizes the nonlinear electro-elasticity theory and the corresponding  linearized incremental theory. 
The nonlinear axisymmetric deformation of a tube is analyzed in Subsec.~\ref{section3.1} for any form of energy density function, and then specialized to the functionally graded Mooney-Rivlin ideal dielectric model in Subsec.~\ref{section3.2}. 
Numerical results are illustrated in Subsec.~\ref{section3.3} for the static nonlinear response and the distributions of biasing fields. Section~\ref{sec4} combines the state-space formalism with the approximate laminate technique. 
We derive the dispersion relations for both T and L waves in Sec.~\ref{section5}. Numerical calculations are presented in Sec.~\ref{Sec6} to validate the effectiveness of the SSM and elucidate the effects of material gradient parameters and biasing fields on the propagation characteristics of T and L waves. 
Finally, we provide a summary in Sec.~\ref{sec7}.


\section{Theoretical background}
\label{section2}

To describe the kinematics of a deformable SEA continuum, the position vector of a given  material point is denoted as $\mathbf{X}$ in the undeformed state (the reference configuration ${{\mathcal{B}}_{r}}$), and as $\mathbf{x}$ in the deformed state (the current configuration ${{\mathcal{B}}_{t}}$). 
The mapping $\mathbf{x}=\bm{\chi} \text{(}\mathbf{X},t\text{)}$, with $t$ being the time variable, is assumed to be sufficiently smooth, and its derivative with respect to $\mathbf{X}$ defines the deformation gradient tensor $\mathbf{F}=\text{Grad}\hskip 1pt\bm{\chi }=\text{Grad}\hskip 1pt\mathbf{x}$, where `$\text{Grad}$' is the gradient operator with respect to ${{\mathcal{B}}_{r}}$. 
The local measure of change in material volume is tracked by $J=\det \mathbf{F}$, which always equals one for an incompressible material, as considered throughout this paper. 
In this work, we adopt the general theoretical framework of nonlinear electro-elasticity and its relevant linearized incremental theory for the superimposed small-amplitude motions developed by \citet{dorfmann2006nonlinear, dorfmann2010electroelastic, dorfmann2014nonlinear}. 
We also show that all seemingly different theories in the literature are actually equivalent and have no substantive difference \citep{wu2016}.


\subsection{Nonlinear electro-elasticity} \label{section2.1}


For incompressible SEA materials, the nonlinear constitutive relations can be written as
\begin{equation}\label{s1-constitutiveL}
\mathbf{T}=\frac{\partial {{\Omega }}}{\partial \mathbf{F}}-p{{\mathbf{F}}^{-1}},\qquad \bm{\mathcal{E}}=\frac{\partial {{\Omega }}}{\partial \bm{\mathcal{D}}},
\end{equation}
where $\Omega = \Omega (\mathbf{F},\bm{\mathcal{D}})$ is the \emph{total energy density function} per unit volume in the reference configuration, $p$ is a Lagrange multiplier related to the incompressibility constraint, $\mathbf{T}={{\mathbf{F}}^{-1}}\bm{\mathbf{\tau }}$ is the \emph{total nominal stress tensor} with $\bm{\mathbf{\tau }}$ being the \emph{total Cauchy stress tensor}, and $\bm{\mathcal{D}}= {{\mathbf{F}}^{-1}} \mathbf{D}$ and $\bm{\mathcal{E}}= {{\mathbf{F}}^{\text{T}}}\mathbf{E}$ are the Lagrangian counterparts of the Eulerian \textit{electric displacement vector} $\mathbf{D}$ and \textit{electric field vector} $\mathbf{E}$, respectively. Note that the superscripts $ (\cdot)^{-1} $ and $(\cdot)^\text{T}$ signify the inverse and transpose of a tensor, respectively.

Accordingly, the corresponding expressions for $\bm{\mathbf{\tau }}$ and $\mathbf{E}$ read as 
\begin{equation}\label{s1-constitutiveE}
\bm{\tau }=\mathbf{F}\frac{\partial {{\Omega }}}{\partial \mathbf{F}}-p\mathbf{I},\qquad \mathbf{E}={{\mathbf{F}}^{-\text{T}}}\frac{\partial {{\Omega }}}{\partial \bm{\mathcal{D}}},
\end{equation}
where $\mathbf{I}$ is the identity tensor. 
For an incompressible and isotropic SEA material, $\Omega$ can be seen as a function of the following five invariants:
\begin{equation} \label{invariants}
I_1=\text{tr}\, \mathbf{C},\quad
I_2=\tfrac{1}{2}{\left[ {{\left( \text{tr}\mathbf{C} \right)}^{2}}-\text{tr}\left( {{\mathbf{C}}^{2}} \right) \right]}, \quad
I_4 = \bm{\mathcal{D} \cdot \mathcal{D}}, \quad
I_5=\bm{\mathcal{D}\cdot \mathbf{C}\mathcal{D}}, \quad 
I_6=\bm{\mathcal{D}\cdot \mathbf{C}}^{2}\bm{\mathcal{D}},
\end{equation}
where $\mathbf{C}={{\mathbf{F}}^{\text{T}}}\mathbf{F}$ is the right Cauchy-Green deformation tensor.

Combination of Eqs.~\eqref{s1-constitutiveE} and \eqref{invariants} gives the total Cauchy stress tensor and the electric field vector as
\begin{equation} \label{initial-constitutive}
\begin{split}
& \bm{\tau }=2\Omega _{1}\mathbf{b}+2\Omega _{2}\left( {{I}_{1}}\mathbf{b}-{{\mathbf{b}}^{2}} \right)-p\mathbf{I}+2\Omega _{5}\mathbf{D}\otimes \mathbf{D}+2\Omega _{6}\left( \mathbf{D}\otimes \mathbf{bD}+\mathbf{bD}\otimes \mathbf{D} \right), \\ 
& \mathbf{E}=2\left( \Omega _{4}{{\mathbf{b}}^{-1}}\mathbf{D}+\Omega _{5}\mathbf{D}+\Omega _{6}\mathbf{bD} \right),
\end{split}
\end{equation}
where $\mathbf{b}={{\mathbf{F}}}\mathbf{F}^{\text{T}}$ is the left Cauchy-Green deformation tensor and the shorthand notation ${{\Omega }_{m}}=\partial \Omega /\partial {{I}_{m}}\text{ }(m=1,2,4,5,6)$ is adopted hereafter.

Under the quasi-electrostatic approximation and in the absence of mechanical body forces as well as free charges and electric currents, the equations of motion, the Gauss law and the Faraday law are
\begin{equation} \label{governingEQ}
\mbox{div\hskip 1pt} \bm{\tau}=\rho \partial^2 \mathbf{x}/\partial t^2, \qquad \mbox{div\hskip 1pt}\mathbf{D}=0,\qquad \mbox{curl\hskip 1pt} \mathbf{E}=\mathbf{0},
\end{equation}
respectively, where $\rho $ is the \emph{unchanged} mass density during the motion, and `$\text{curl}$' and `$\text{div}$' are the curl and divergence operators in ${{\mathcal{B}}_{t}}$, respectively. 
We emphasize that $\bm{\mathbf{\tau }}$ takes account of the contribution of the electric body forces and that the conservation of angular momentum ensures the symmetry of $\bm{\tau }$.

The mechanical and electric boundary conditions to be satisfied on the boundary $\partial {{\mathcal{B}}_{t}}$ are expressed in Eulerian form as
\begin{equation} \label{boundary}
\bm{\tau}^\text{T} \bm{n}_t={{\mathbf{t}}^{\text{a}}}, \qquad  \mathbf{E}\times \mathbf{n}_t= \mathbf{0},\qquad \mathbf{D \cdot n}_t=-{{\sigma }_{\text{f}}},
\end{equation}
where ${{\mathbf{n}}_{t}}$ is the outward unit normal vector of the current configuration ${{\mathcal{B}}_{t}}$, and ${{\mathbf{t}}^{\text{a}}}$ and ${{\sigma }_{\text{f}}}$ denote the applied mechanical traction vector per unit area of $\partial {{\mathcal{B}}_{t}}$ and the free surface charge density on $\partial {{\mathcal{B}}_{t}}$, respectively. 
Note that, since an electric voltage will be applied to the surfaces of the SEA body coated with flexible electrodes, the electric field in the surrounding vacuum can been neglected in Eq.~\eqref{boundary}.


\subsection{Linearized theory for incremental motions}\label{Sec2-2}


When  a time-dependent infinitesimal incremental motion $\mathbf{\dot{x}}(\mathbf{X},t)$ is superimposed on a finitely deformed SEA body that occupies a static configuration $\mathcal{B}$ associated with the mapping $\mathbf{x}=\bm{\chi} \text{(}\mathbf{X})$, the linearized incremental incompressibility condition, governing equations and constitutive laws for incompressible SEA materials can be written in \emph{updated Lagrangian} form as
\begin{align}
& \mbox{div\hskip 1pt}\mathbf{u} =\mbox{tr\hskip 1pt}\mathbf{H}=0,
\label{incompre}
\\
&
\mbox{div\hskip 1pt}{{\mathbf{\dot{T}}}_{0}}=\rho \partial^2 {\mathbf{u}}/\partial t^2, \qquad \mbox{div\hskip 1pt}{{\bm{\dot{\mathcal{D}}}}_{0}}=0,\qquad \mbox{curl\hskip 1pt}{{\bm{\dot{\mathcal{E}}}}_{0}}=\mathbf{0},  \label{incre-governEQ}
\end{align}
where
\begin{equation} \label{increconsti}
{{\mathbf{\dot{T}}}_{0}}={{\bm{\mathcal{A}}}_{0}}\mathbf{H}+{{\bm{\mathcal{M}}}_{0}}{{\bm{\dot{\mathcal{D}}}}_{0}}+p\mathbf{H}-\dot{p}\mathbf{I},\qquad {{\bm{\dot{\mathcal{E}}}}_{0}}=\bm{\mathcal{M}}_{0}^{\text{T}}\mathbf{H}+{{\bm{\mathcal{R}}}_{0}}{{\bm{\dot{\mathcal{D}}}}_{0}}.
\end{equation}
Here a superposed dot indicates an increment in the quantity concerned, $\mathbf{u}(\mathbf{x},t)= \mathbf{\dot{x}}(\mathbf{X},t)$ is the incremental mechanical displacement vector, $\mathbf{H}=\mbox{grad\hskip 1pt}\mathbf{u}$ denotes the incremental displacement gradient tensor with `$\text{grad}$' being the gradient operator in $\mathcal{B}$, $\dot{p}$ is the incremental Lagrange multiplier, and ${{\mathbf{\dot{T}}}_{0}}$, ${{\bm{\dot{\mathcal{D}}}}_{0}}$ and ${{\bm{\dot{\mathcal{E}}}}_{0}}$ represent the \emph{push-forward} versions of the corresponding Lagrangian increments, all updating the reference configuration from the original unstressed reference configuration ${{\mathcal{B}}_{r}}$ to the initial deformed configuration $\mathcal{B}$. Note that a subscript `0' is utilized to identify the resultant push-forward variables.

In component notation, the components of the \emph{instantaneous} electro-elastic moduli tensors ${{\bm{\mathcal{A}}_{0}}}$, ${{\bm{\mathcal{M}}_{0}}}$ and ${{\bm{\mathcal{R}}_{0}}}$ in Eq.~\eqref{increconsti} are defined as
\begin{equation}\label{s1-effective-material}
{{\mathcal{A}}_{0piqj}}={{F}_{p\alpha }}{{F}_{q\beta }}{{\mathcal{A}}_{\alpha i\beta j}},\qquad 
{{\mathcal{R}}_{0ij}}=F_{\alpha i}^{-1}F_{\beta j}^{-1}{{\mathcal{R}}_{\alpha \beta }},\qquad
{{\mathcal{M}}_{0piq}}={{F}_{p\alpha }}F_{\beta q}^{-1}{{\mathcal{M}}_{\alpha i\beta }},
\end{equation}
where  $\bm{\mathcal{A}}={{\partial }^{2}}\Omega /(\partial \mathbf{{F}}\partial \mathbf{{F}})$, $\bm{\mathcal{M}}={{\partial }^{2}}\Omega /(\partial \mathbf{{F}}\partial \bm{\mathcal{{D}}})$ and $\bm{\mathcal{R}}={{\partial }^{2}}\Omega /(\partial \bm{\mathcal{{D}}}\partial \bm{\mathcal{{D}}})$ are the \emph{referential} electro-elastic moduli tensors.
Note the following symmetries
\begin{equation}
{{\mathcal{A}}_{0piqj}}={{\mathcal{A}}_{0qjpi}},\qquad {{\mathcal{R}}_{0ij}}={{\mathcal{R}}_{0ji}},\qquad
\mathcal{M}_{0piq} = \mathcal{M}_{0ipq},
\end{equation}

When we neglect the increments of electrical variables in vacuum, the updated Lagrangian incremental forms of the mechanical and electric boundary conditions satisfied on $\partial \mathcal{B}$ are
\begin{equation}\label{s1-incremental-boundary}
\mathbf{\dot{T}}_{0}^{\text{T}}\mathbf{n}=\mathbf{\dot{t}}_{0}^{\text{A}}, \qquad {{\bm{\dot{\mathcal{E}}}}_{0}}\times \mathbf{n}=\mathbf{0},\qquad {{\bm{\dot{\mathcal{D}}}}_{0}}\cdot \mathbf{n}=-{{\dot{\sigma }}_{\text{F0}}},
\end{equation}
where ${{\mathbf{n}}}$ is the outward unit normal vector of the static configuration ${{\mathcal{B}}}$, $\mathbf{\dot{t}}_{0}^{\text{A}}$ is the updated Lagrangian incremental traction vector per unit area of the boundary $\partial \mathcal{B}$, and ${{\dot{\sigma }}_{\text{F0}}}$ is the incremental surface charge density on $\partial \mathcal{B}$. 
When a hydrostatic pressure $P_{\text{a}}$ is applied to the boundary $\partial \mathcal{B}$, the incremental mechanical boundary condition {\color{black}  Eq.~\eqref{s1-incremental-boundary}$_1$} becomes \citep{ottenio2007acoustic, bustamante2013axisymmetric, wu2018propagation}
\begin{equation} \label{Pcondition}
\mathbf{\dot{T}}_{0}^{\text{T}}\mathbf{n}={{P}_{\text{a}}}{{\mathbf{H}}^{\operatorname{T}}}\mathbf{n}-{{\dot{P}}_{\text{a}}}\mathbf{n},
\end{equation}
where $\dot{P}_{\text{a}}$ denotes the increment of the applied pressure.


\section{Axisymmetric deformation of a functionally graded SEA tube} \label{section3}


The nonlinear axisymmetric deformations of a \emph{homogeneous} SEA tube subject to different electro-mechanical biasing fields have been examined by several authors \citep{dorfmann2006nonlinear, zhu2010large, shmuel2013axisymmetric, melnikov2016finite, wu2017guided}.
For a \emph{functionally graded} hyperelastic hollow cylinder without electro-mechanical coupling, several studies have also been carried out on the static nonlinear response \citep{batra2009inflation, chen2017bifurcation, wu2018propagation}. 
This section further extends these results to consider the finite axisymmetric deformations of \emph{functionally graded SEA tubes} under the combined actions of axial pre-stretch, radial electric voltage and pressure difference.


\subsection{General dielectric material model}\label{section3.1}


We consider an incompressible isotropic functionally graded SEA tube coated with flexible electrodes on its inner and outer surfaces. 
Figs.~\ref{Fig1}(a)-(b) show schematic diagrams of its nonlinear axisymmetric deformation under electro-mechanical activation. 

The cylindrical coordinates in the undeformed and deformed configurations are denoted by $(R,\mit{\Theta},Z)$ and $\left( r,\theta ,z \right)$, respectively, with corresponding vector bases $(\mathbf E_R, \mathbf E_\Theta, \mathbf E_Z)$ and $(\mathbf e_r, \mathbf e_\theta, \mathbf e_z)$, respectively. 
Let the length, inner and outer radii of the tube in the undeformed configuration be $L$, $A$ and $B$, respectively, with the thickness $H=B-A$. 
The tube is deformed by an axial stretching, a pressure difference and an electric voltage in the radial direction to maintain the axial symmetry. The current length, inner and outer radii are $\ell$, $a$ and $b$, respectively, with the thickness $h=b-a$.
 
We take the nonlinear axisymmetric deformation in the following form:
\begin{equation} \label{kinematic}
r=\sqrt{G+\lambda_z^{-1}R^2},\qquad \theta =\Theta ,\qquad z={{\lambda }_{z}}Z,
\end{equation}
where $G={a}^{2}-\lambda _{z}^{-1}A^2$ and ${\lambda }_{z}$ is the uniform axial principal stretch. 
Note that it is a simple exercise to check that this deformation is isochoric and thus compatible with the incompressibility constraint. 
The corresponding deformation gradient has components $\mathbf{F}=\text{diag}[\lambda_r, {{\lambda }_{\theta }},{{\lambda }_{z}}]$ in the $\mathbf e_i \otimes \mathbf E_k$ basis, where ${{\lambda }_{r}}=\lambda _{\theta }^{-1}\lambda _{z}^{-1}$ and ${{\lambda }_{\theta }}=r/R$ are the radial and circumferential principal stretches, respectively. 

The underlying Eulerian electric displacement vector is taken to be radial only, so that $\mathbf{D}=[D_r, 0,0]^{\text{T}}$, say, and its Lagrangian counterpart is $\bm{\mathcal{D}}={{\mathbf{F}}^{-1}}\mathbf{D} =[\mathcal D_r, 0,0]^{\text{T}}$, where ${{\mathcal{D}}_{r}}={{\lambda }_{\theta }}{{\lambda }_{z}}{{D}_{r}}$. 

From Eq.~\eqref{kinematic}$_1$, we obtain the connection between the circumferential stretches ${\lambda }_{a}=a/A$  on the inner surface   and ${\lambda }_{b}=b/B$ on the outer surface, as
\begin{equation} \label{lamdaab}
\lambda _{a}^{2}{{\lambda }_{z}}-1={{\Lambda }^{2}}\left( \lambda _{\theta }^{2}{{\lambda }_{z}}-1 \right)=\eta^{2}\left( \lambda _{b}^{2}{{\lambda }_{z}}-1 \right),
\end{equation}
where ${{\eta }}=B/A$ is the outer-to-inner radius ratio, and $\Lambda =R/A\in [1,\eta ]$ is the dimensionless radial coordinate in the undeformed configuration.

In terms of the stretches and the radial electric displacement, the five independent invariants in Eq.~\eqref{invariants} and the nonzero components of $\bm{\tau}$ and $\mathbf{E}$ in Eq.~\eqref{initial-constitutive} are calculated as
\begin{align} \label{invariant-new}
{{I}_{1}}&=\lambda _{\theta }^{-2}\lambda _{z}^{-2}+\lambda _{\theta }^{2}+\lambda _{z}^{2},\qquad {{I}_{2}}=\lambda _{\theta }^{2}\lambda _{z}^{2}+\lambda _{\theta }^{-2}+\lambda _{z}^{-2}, \notag\\
{{I}_{4}}&=\lambda _{\theta }^{2}\lambda _{z}^{2}D_{r}^{2},\qquad {{I}_{5}}=\lambda _{\theta }^{-2}\lambda _{z}^{-2}{{I}_{4}},\qquad {{I}_{6}}=\lambda _{\theta }^{-4}\lambda _{z}^{-4}{{I}_{4}},
\end{align}
and 
\begin{align} \label{iniconstitutive}
& {{\tau }_{rr}}=2\lambda _{\theta }^{-2}\lambda _{z}^{-2}\left[ {{\Omega }_{1}}+{{\Omega }_{2}}(\lambda _{\theta }^{2}+\lambda _{z}^{2}) \right]+2\left( {{\Omega }_{5}}+2{{\Omega }_{6}}\lambda _{\theta }^{-2}\lambda _{z}^{-2} \right)D_{r}^{2}-p, \notag\\ 
& {{\tau }_{\theta \theta }}=2\lambda _{\theta }^{2}\left[ {{\Omega }_{1}}+{{\Omega }_{2}}(\lambda _{\theta }^{-2}\lambda _{z}^{-2}+\lambda _{z}^{2}) \right]-p, \notag\\ 
& {{\tau }_{zz}}=2\lambda _{z}^{2}\left[ {{\Omega }_{1}}+{{\Omega }_{2}}(\lambda _{\theta }^{-2}\lambda _{z}^{-2}+\lambda _{\theta }^{2}) \right]-p, \notag\\ 
& {{E}_{r}}=2({{\Omega }_{4}}\lambda _{\theta }^{2}\lambda _{z}^{2}+{{\Omega }_{5}}+{{\Omega }_{6}}\lambda _{\theta }^{-2}\lambda _{z}^{-2}){{D}_{r}}.
\end{align}

Now by defining a reduced energy density function $\omega^*$ as ${{\omega }^{*}}({{\lambda }_{\theta }},{{\lambda }_{z}},{{I}_{4}})=\Omega ({{I}_{1}},{{I}_{2}},{{I}_{4}},{{I}_{5}},{{I}_{6}})$, we obtain the following relations from Eqs.~\eqref{invariant-new} and \eqref{iniconstitutive} for a general dielectric material model \citep{melnikov2016finite, wu2017guided}:
\begin{equation} \label{universal}
{{\lambda }_{\theta }}\omega _{{{\lambda }_{\theta }}}^{*}={{\tau }_{\theta \theta }}-{{\tau }_{rr}},\qquad {{\lambda }_{z}}\omega _{{{\lambda }_{z}}}^{*}={{\tau }_{zz}}-{{\tau }_{rr}},\qquad {{E}_{r}}=2\lambda _{\theta }^{2}\lambda _{z}^{2}\omega _{4}^{*}{{D}_{r}},
\end{equation}
where $\omega _{{{\lambda }_{\theta }}}^{*}=\partial {{\omega }^{*}}/\partial {{\lambda }_{\theta }}$, $\omega _{{{\lambda }_{z}}}^{*}=\partial {{\omega }^{*}}/\partial {{\lambda }_{z}}$ and $\omega _{4}^{*}=\partial {{\omega }^{*}}/\partial {{I}_{4}}$.

Due to the considered axisymmetric deformation, all initial physical quantities depend only on $r$. Therefore, Faraday's law {\color{black}  Eq.~\eqref{governingEQ}$_3$} is satisfied automatically, and  Gauss's law {\color{black}  Eq.~\eqref{governingEQ}$_2$} and the equilibrium equation $\text{div}\; \bm{\tau }=\mathbf{0}$ simplify to
\begin{equation}\label{equi}
\frac{1}{r}\frac{\partial (r{{D}_{r}})}{\partial r}=0,\qquad
\frac{\text{d}{{\tau }_{rr}}}{\text{d}r}=\frac{{{\tau }_{\theta \theta }}-{{\tau }_{rr}}}{r}=\frac{{{\lambda }_{\theta }}\omega _{{{\lambda }_{\theta }}}^{*}}{r},
\end{equation}
respectively, where Eq.~\eqref{universal}$_1$ has been used. Integration of Eq.~\eqref{equi}$_1$ leads to
\begin{equation} \label{electricDIS}
{{D}_{r}}=\frac{Q(a)}{2\pi r{{\lambda }_{z}}L}=-\frac{Q(b)}{2\pi r{{\lambda }_{z}}L},
\end{equation}
where $Q(a)$ and $Q(b)$ (satisfying  $Q(a)+Q(b)=0$) are total free surface charges on the inner and outer surfaces of the deformed tube, respectively.
These charges are related to the free surface charge densities ${{\sigma}_{\text{fa}}}$ and ${{\sigma}_{\text{fb}}}$ on the inner and outer deformed surfaces as ${{\sigma}_\text{fa}}={Q(a)}/({2\pi a{{\lambda }_{z}}L})$ and ${{\sigma}_{\text{fb}}}={Q(b)}/({2\pi b{{\lambda }_{z}}L})$, respectively. 
Note that the initial boundary condition {\color{black}  Eq.~\eqref{boundary}$_3$} (i.e., ${{D}_{r}}(a)={{\sigma }_{\text{fa}}}$ and ${{D}_{r}}(b)=-{{\sigma }_{\text{fb}}}$) has been employed to derive Eq.~\eqref{electricDIS}.

Moreover, the curl-free electric field can be expressed as $\mathbf{E}=-\text{grad}\phi $ by introducing an electrostatic potential $\phi $, which here yields the only nonzero electric field component as ${{E}_{r}}=-\text{d}\phi /\text{d}r$. Inserting Eq.~\eqref{electricDIS} into Eq.~\eqref{universal}$_3$ and integrating the resultant equation from the inner surface to the outer one, we obtain
\begin{equation} \label{voltage-Q}
V={{\lambda }_{z}}\frac{Q(a)}{\pi L}\int_{a}^{b}{\lambda _{\theta }^{2}\omega _{4}^{*}\frac{\text{d}r}{r}},
\end{equation}
where $V=\phi (a)-\phi (b)$ is the electric potential difference or \emph{electric voltage} between the inner and outer surfaces. Thus, Eq.~\eqref{voltage-Q} represents a general relationship between the electric voltage $V$ and the surface free charge $Q$, which is affected by the initial deformation.

Additionally, integrating Eq.~\eqref{equi}$_2$ from $a$ to $b$ and assuming that the internal and external pressures $P_{\text{inn}}$ and $P_{\text{out}}$ are applied to the inner and outer surfaces (i.e., ${{\tau }_{rr}}(a)=-{{P}_{\text{inn}}}$ and ${{\tau }_{rr}}(b)=-{{P}_{\text{out}}}$), we have
\begin{equation} \label{POmega}
{{P}_{\text{out}}}-{{P}_{\text{inn}}}=-\int_{a}^{b}{{{\lambda }_{\theta }}\omega _{{{\lambda }_{\theta }}}^{*}}\frac{\text{d}r}{r},
\end{equation}
Note that ${\lambda}_{b}$ may be expressed in terms of ${\lambda }_{a}$ and ${\lambda }_{z}$ by Eq.~\eqref{lamdaab}. Thus, Eq.~\eqref{POmega} establishes a general nonlinear relation between the pressure difference (or the net pressure) $\Delta P={{P}_{\text{out}}}-{{P}_{\text{inn}}}$, the electrical variable $Q$ or $V$ (included in ${{\omega^* }}$) and the inner radius $a$ (measured by ${{\lambda }_{a}}$) for any given geometric parameter $\eta$ and the axial pre-stretch ${\lambda }_{z}$. Similarly, the radial normal stress is obtained by integrating Eq.~\eqref{equi}$_2$ from $a$ to $r$ as 
\begin{equation} \label{sigmarr}
{{\tau }_{rr}}(r)=\int_{a}^{r}{{{\lambda }_{\theta }}\omega _{{{\lambda }_{\theta }}}^{*}}\frac{\text{d}r}{r}-{{P}_{\text{inn}}}.
\end{equation}

In order to maintain the fixed axial stretch and the axisymmetric deformation state, an axial load is required at the ends of the tube, which can be open or closed. Using Eq.~\eqref{universal}$_{1,2}$, the equilibrium equation \eqref{equi}$_2$ and the initial mechanical boundary conditions, the axial normal stress ${{\tau }_{zz}}$ and the \emph{resultant} axial force $N$ are written as
\begin{equation} \label{sigmazz}
{{\tau }_{zz}}=\frac{1}{2}\left[ \frac{1}{r}\frac{\text{d}}{\text{d}r}\left( {{r}^{2}}{{\tau }_{rr}} \right)-{{\lambda }_{\theta }}\omega _{{{\lambda }_{\theta }}}^{*} \right]+{{\lambda }_{z}}\omega _{{{\lambda }_{z}}}^{*},
\end{equation}
and
\begin{equation} \label{axial-forceTotal}
N=2\pi \int_{a}^{b}{{{\tau }_{zz}}(r)r\text{d}r}=\pi \int_{a}^{b}{\left( 2{{\lambda }_{z}}\omega _{{{\lambda }_{z}}}^{*}-{{\lambda }_{\theta }}\omega _{{{\lambda }_{\theta }}}^{*} \right)r\text{d}r}+\pi \left( {{a}^{2}}{{P}_{\text{inn}}}-{{b}^{2}}{{P}_{\text{out}}} \right),
\end{equation}
which is required in the case where the tube has \emph{open ends}. 
However, in the case of \emph{closed ends}, the resultant axial force $N$ includes a contribution from the internal and external pressures on the tube ends. Consequently, a \emph{reduced} axial force (i.e. the externally applied axial force) is defined as
\begin{equation} \label{axial-forceRed}
{{N}_{r}}\equiv N-\pi \left( {{a}^{2}}{{P}_{\text{inn}}}-{{b}^{2}}{{P}_{\text{out}}} \right)=\pi \int_{a}^{b}{\left( 2{{\lambda }_{z}}\omega _{{{\lambda }_{z}}}^{*}-{{\lambda }_{\theta }}\omega _{{{\lambda }_{\theta }}}^{*} \right)r\text{d}r},
\end{equation}
which eliminates the contribution of the pressures from the resultant axial force $N$.

We emphasize that the theoretical formulations obtained above are completely general for an incompressible isotropic functionally graded SEA tube characterized by an arbitrary reduced energy function $\omega^*$. The integrations {\color{black} in Eqs.~\eqref{voltage-Q}-\eqref{axial-forceRed}}
can be conducted analytically or numerically once the form of $\omega^*$ is prescribed.


\subsection{Functionally graded  Mooney-Rivlin ideal dielectric model}\label{section3.2}


For definiteness, the previous results are now specialized to the generalized functionally graded Mooney-Rivlin ideal dielectric model, which is characterized by the following energy density function:
\begin{equation} \label{MR1}
\Omega ={{\Omega }_{0}}\left( {{I}_{1}},{{I}_{2}} \right)+\frac{1}{2\varepsilon \left( R \right)} I_5,\qquad
{{\Omega }_{0}}\left( {{I}_{1}},{{I}_{2}} \right)=\frac{\mu_1(R)}{2}(I_1 - 3) - \frac{\mu_2(R)}{2}(I_2 - 3),
\end{equation}
where ${{\mu }_{1}}(R)$, ${{\mu }_{2}}(R)$ and $\varepsilon(R)$ are the electro-mechanical material parameters, with properties depending on the undeformed radial coordinate $R$. 
The quantity ${{\mu }_{1}}(R)-{{\mu }_{2}}(R)$ is the shear modulus ${{\mu }}(R)>0$ in the absence of electrical fields.
In view of Eq.~\eqref{invariant-new}$_{1,2,4}$, the reduced form of the energy density is
\begin{align}  \label{MR2}
& {{\omega }^{*}}({{\lambda }_{\theta }},{{\lambda }_{z}},{{I}_{4}})=\omega _{0}^{*}\left( {{\lambda }_{\theta }},{{\lambda }_{z}} \right)+\frac{\lambda _{\theta }^{-2}\lambda _{z}^{-2}{{I}_{4}}}{2\varepsilon \left( R \right)}, \notag\\ 
& \omega _{0}^{*}\left( {{\lambda }_{\theta }},{{\lambda }_{z}} \right)=\frac{{{\mu }_{1}}(R)}{2}(\lambda _{\theta }^{-2}\lambda _{z}^{-2}+\lambda _{\theta }^{2}+\lambda _{z}^{2}-3) -\frac{{{\mu }_{2}}(R)}{2}(\lambda _{\theta }^{2}\lambda _{z}^{2}+\lambda _{\theta }^{-2}+\lambda _{z}^{-2}-3).
\end{align}

To be specific, we assume \emph{affine variations} of material parameters \citep{batra2009inflation, wu2018propagation}, as follows
\begin{align}  \label{gradientPara}
& {{\mu }_{1}}(R)={{\mu }_{10}}\left( 1+{{\beta }_{1}}\Lambda  \right), \qquad {{\mu }_{2}}(R)={{\mu }_{20}}\left( 1+{{\beta }_{2}}\Lambda  \right), \qquad \varepsilon (R)={{\varepsilon }_{a0}}\left( 1+{{\beta }_{3}}\Lambda  \right),
\end{align}
where ${{\mu }_{10}}$ and ${{\mu }_{20}}$ are  elastic material constants (in N/m$^2$), $\varepsilon_{a0}$ is a dielectric constant (in F/m), $\beta_1$ and $\beta_2$ are  \emph{elastic moduli gradient parameters}, and $\beta_3$ is the \emph{permittivity gradient parameter.} Note that $\beta_1$, $\beta_2$, $\beta_3$ are dimensionless and characterize the functionally graded properties of the SEA tube.

Differentiation of Eq.~\eqref{MR2}$_1$ with respect to $I_4$, $\lambda _{\theta }$ and $\lambda _{z }$ yields in turn 
\begin{equation} \label{differentiation}
\lambda _{\theta }^{2}\omega _{4}^{*}=\frac{\lambda _{z}^{-2}}{2\varepsilon \left( R \right)},
\qquad 
{{\lambda }_{\theta }}\omega _{{{\lambda }_{\theta }}}^{*}={{\lambda }_{\theta }}\omega _{0,{{\lambda }_{\theta }}}^{*}-\frac{D_{r}^{2}}{\varepsilon \left( R \right)}
\qquad 
{{\lambda }_{z}}\omega _{{{\lambda }_{z}}}^{*}={{\lambda }_{z}}\omega _{0,{{\lambda }_{z}}}^{*}-\frac{D_{r}^{2}}{\varepsilon \left( R \right)},
\end{equation}
where $\omega _{0,{{\lambda }_{\theta }}}^{*}=\partial \omega _{0}^{*}/\partial {{\lambda }_{\theta }}$, $\omega _{0,{{\lambda }_{z}}}^{*}=\partial \omega _{0}^{*}/\partial {{\lambda }_{z}}$ and Eq.~\eqref{invariant-new}$_3$ has been used.
Substituting Eqs.~\eqref{differentiation}$_1$ and \eqref{gradientPara}$_3$ into Eq.~\eqref{voltage-Q} and conducting the integration, we obtain
the \emph{dimensionless electric voltage} ${{V}^{*}}=V\sqrt{{{\varepsilon }_{a0}}/{{\mu }_{10}}}/H$ as
\begin{equation} \label{voltage-Q1}
{{V}^{*}}=\frac{{{Q}^{*}}}{2{{\lambda }_{z}}\left( {{\eta }}-1 \right)\left( 1+{{G}^{*}}\beta _{3}^{2}{{\lambda }_{z}} \right)}\left. \left[ V_{0}^{*}\left( \Lambda  \right) \right] \right|_{1}^{{{\eta }}},
\end{equation} 
where  ${{Q}^{*}}=Q(a)/(2\pi AL\sqrt{{{\mu }_{10}}{{\varepsilon }_{a0}}})$ is the \emph{dimensionless surface charge}, ${{G}^{*}}=G/A^2=\lambda _{a}^{2}-\lambda _{z}^{-1}$, and $\left. \left[ V_{0}^{*}\left( \Lambda  \right) \right] \right|_{1}^{{\eta }}=V_{0}^{*}\left( {\eta} \right)-V_{0}^{*}\left( 1 \right)$ with
\begin{equation} \label{voltage*}
V_{0}^{*}\left( \Lambda  \right)=\ln \frac{{{\lambda }_{z}}{{G}^{*}}+{{\Lambda }^{2}}}{{{\left( 1+{{\beta }_{3}}\Lambda  \right)}^{2}}}+2{{\beta }_{3}}\sqrt{{{\lambda }_{z}}{{G}^{*}}}\hskip 1pt \text{tan}^{-1}\frac{\Lambda }{\sqrt{{{\lambda }_{z}}{{G}^{*}}}}.
\end{equation}
In the absence of functional gradients (i.e., $\beta_i=0$), the relation \eqref{voltage-Q1} recovers Eq.~(51) in the paper by \citet{melnikov2016finite} and Eq.~(63)$_2$ in the paper by \citet{wu2017guided} for  homogeneous SEA tubes.

After substituting Eq.~\eqref{differentiation}$_{2,3}$ into Eqs.~\eqref{POmega} and \eqref{axial-forceRed}, the pressure difference and the reduced axial force become
\begin{align} \label{PandN}
& \Delta P={{P}_{\text{out}}}-{{P}_{\text{inn}}}=-\int_{a}^{b}{{{\lambda }_{\theta }}\omega _{0,{{\lambda }_{\theta }}}^{*}}\frac{\text{d}r}{r}+\int_{a}^{b}{\frac{D_{r}^{2}}{\varepsilon \left( R \right)}}\frac{\text{d}r}{r}, \notag\\ 
& {{N}_{r}}=\pi \int_{a}^{b}{\left( 2{{\lambda }_{z}}\omega _{0,{{\lambda }_{z}}}^{*}-{{\lambda }_{\theta }}\omega _{0,{{\lambda }_{\theta }}}^{*} \right)r\text{d}r}-\pi \int_{a}^{b}{\frac{D_{r}^{2}}{\varepsilon \left( R \right)}r\text{d}r}.
\end{align}
Inserting Eq.~\eqref{electricDIS} into the \emph{second} terms of Eqs.~\eqref{PandN}$_{1}$ and \eqref{PandN}$_{2}$ and integrating the resultant expressions, we have
\begin{multline} \label{SecondpartP}
\int_{a}^{b}{\frac{D_{r}^{2}}{\varepsilon \left( R \right)}}\frac{\text{d}r}{r}={{\left( \frac{Q(a)}{2\pi {{\lambda }_{z}}L} \right)}^{2}}\frac{A{{\lambda }_{z}}}{2{{\varepsilon }_{a0}}{{\left( {{A}^{2}}+G\beta _{3}^{2}{{\lambda }_{z}} \right)}^{2}}}\left[ A\beta _{3}^{2}\ln \frac{{{\lambda }_{z}}G+{{R}^{2}}}{{{\left( A+{{\beta }_{3}}R \right)}^{2}}} \right. \\ 
 \left. \left. +\frac{{{\beta }_{3}}\left( G\beta _{3}^{2}{{\lambda }_{z}}-{{A}^{2}} \right)}{\sqrt{{{\lambda }_{z}}G}}\text{tan}^{-1}\frac{R}{\sqrt{{{\lambda }_{z}}G}}+\left( {{A}^{2}}+G\beta _{3}^{2}{{\lambda }_{z}} \right)\frac{{{\beta }_{3}}R-A}{{{\lambda }_{z}}G+{{R}^{2}}} \right] \right|_{A}^{B},
\end{multline}
and
\begin{multline} \label{SecondpartNr}
\pi \int_{a}^{b}{\frac{D_{r}^{2}}{\varepsilon \left( R \right)}r\text{d}r}={{\left( \frac{Q(a)}{2\pi {{\lambda }_{z}}L} \right)}^{2}}\frac{\pi A}{2{{\varepsilon }_{a0}}\left( {{A}^{2}}+\beta _{3}^{2}{{\lambda }_{z}}G \right)}\left[ A\ln \frac{{{\lambda }_{z}}G+{{R}^{2}}}{{{\left( A+{{\beta }_{3}}R \right)}^{2}}} \right. \\ 
\left. \left. +2{{\beta }_{3}}\sqrt{{{\lambda }_{z}}G}\hskip 2pt\text{tan}^{-1}\frac{R}{\sqrt{{{\lambda }_{z}}G}} \right] \right|_{A}^{B}.
\end{multline}
In addition, differentiating Eq.~\eqref{MR2}$_2$ with respect to $\lambda _{\theta }$ and $\lambda _{z }$ leads to
\begin{align} \label{differentiation1}
{{\lambda }_{\theta }}\omega _{0,{{\lambda }_{\theta }}}^{*}=\left[ {{\mu }_{1}}(R)-{{\mu }_{2}}(R)\lambda _{z}^{2} \right]\frac{\lambda _{z}^{2}{{r}^{4}}-{{R}^{4}}}{\lambda _{z}^{2}{{R}^{2}}{{r}^{2}}}, \qquad {{\lambda }_{z}}\omega _{0,{{\lambda }_{z}}}^{*}=\left[ \frac{{{\mu }_{1}}(R)}{{{r}^{2}}}-\frac{{{\mu }_{2}}(R)}{{{R}^{2}}} \right]\frac{\lambda _{z}^{4}{{r}^{2}}-{{R}^{2}}}{\lambda _{z}^{2}},  
\end{align}
where we have used ${{\lambda }_{\theta }}=r/R$. 
Thus, substituting Eq.~\eqref{differentiation1} into the \emph{first} terms of Eq.~\eqref{PandN}$_{1,2}$ and carrying out the integration of the resultant expressions, we obtain
\begin{multline} \label{FirstpartP}
-\int_{a}^{b}{{{\lambda }_{\theta }}\omega _{0,{{\lambda }_{\theta }}}^{*}}\frac{\text{d}r}{r}  =\left[ \left( \frac{{{\mu }_{10}}}{{{\lambda }_{z}}}-{{\mu }_{20}}{{\lambda }_{z}} \right)\left( \ln {{\lambda }_{\theta }}+\frac{G}{2{{r}^{2}}} \right) \right. \\ 
 \left. \left. +\frac{1}{2A}\left( \frac{{{\mu }_{10}}}{{{\lambda }_{z}}}{{\beta }_{\text{1}}}-{{\mu }_{20}}{{\lambda }_{z}}{{\beta }_{\text{2}}} \right)\left( \frac{GR}{{{r}^{2}}}-3\sqrt{{{\lambda }_{z}}G} \hskip 2pt {{\tan }^{-1}}\frac{R}{\sqrt{{{\lambda }_{z}}G}} \right) \right] \right|_{A}^{B},  
\end{multline}
and
\begin{multline} \label{FirstpartNr}
\pi \int_{a}^{b}{\left( 2{{\lambda }_{z}}\omega _{0,{{\lambda }_{z}}}^{*} - {{\lambda }_{\theta }}\omega _{0,{{\lambda }_{\theta }}}^{*} \right) r\text{d}r} = \pi \left[ \left( {{\mu }_{10}}{{\lambda }_{z}}+\frac{{{\mu }_{20}}}{\lambda _{z}^{3}} \right){{R}^{2}} \right. \\ 
- \left( \frac{{{\mu }_{10}}}{{{\lambda }_{z}}} + {{\mu }_{20}}{{\lambda }_{z}} \right) \left( \frac{{{R}^{2}}}{{{\lambda }_{z}}} - G\ln {{\lambda }_{\theta }} \right)  + \frac{2{{R}^{3}}}{3A}\left( {{\mu }_{10}}{{\lambda }_{z}}{{\beta }_{1}} + \frac{{{\mu }_{20}}{{\beta }_{2}}}{\lambda _{z}^{3}} \right) \\
- \left. \left. \left( \frac{{{\mu }_{10}}{{\beta }_{\text{1}}}}{A{{\lambda }_{z}}}+\frac{{{\mu }_{20}}{{\lambda }_{z}}{{\beta }_{\text{2}}}}{A} \right)\left( \frac{2{{R}^{3}}}{3{{\lambda }_{z}}}+G\sqrt{{{\lambda }_{z}}G}\hskip 2pt {{\text{tan}}^{-1}}\frac{R}{\sqrt{{{\lambda }_{z}}G}} \right) \right] \right|_{A}^{B}.
\end{multline}

Consequently, the combination of Eqs.~\eqref{SecondpartP}-\eqref{SecondpartNr} and \eqref{FirstpartP}-\eqref{FirstpartNr} yields the dimensionless pressure difference $\Delta {{P}^{*}}=\Delta P/{{\mu }_{10}}$ and the \textit{reduced} axial force $N_{r}^{*}={{N}_{r}}/(\pi {{\mu }_{10}}{{A}^{2}})$ as
\begin{align} \label{PandNr}
& \Delta {{P}^{*}}=\left. \left[ \Delta P_{1}^{*}\left( \Lambda  \right)+\Delta P_{2}^{*}\left( \Lambda  \right) \right] \right|_{1}^{{{\eta }}}, \notag\\ & N_{r}^{*}={{N}^{*}}-\left( \lambda _{a}^{2}P_{\text{inn}}^{*}-\lambda _{b}^{2}\eta^{2} P_{\text{out}}^{*} \right)=\left. \left[ N_{r1}^{*}\left( \Lambda  \right)-N_{r2}^{*}\left( \Lambda  \right) \right] \right|_{1}^{{{\eta }}},
\end{align}
where ${{N}^{*}}=N/(\pi {{\mu }_{10}}{{A}^{2}})$ is the dimensionless \textit{resultant} axial force, and $P_{\text{inn}}^{*}={{P}_{\text{inn}}}/{{\mu }_{10}}$,  $P_{\text{out}}^{*}={{P}_{\text{out}}}/{{\mu }_{10}}$ denote the dimensionless internal and external pressures, respectively, and
\begin{align} \label{diffparts}
\Delta P_{1}^{*}\left( \Lambda  \right) & =\left( \frac{1}{{{\lambda }_{z}}}-\frac{{{\mu }_{20}}}{{{\mu }_{10}}}{{\lambda }_{z}} \right)\left( \ln {{\lambda }_{\theta }}+\frac{{{G}^{*}}}{2\left( {{G}^{*}}+\lambda _{z}^{-1}{{\Lambda }^{2}} \right)} \right) \notag\\ 
& +\frac{1}{2}\left( \frac{{{\beta }_{\text{1}}}}{{{\lambda }_{z}}}-\frac{{{\mu }_{20}}}{{{\mu }_{10}}}{{\lambda }_{z}}{{\beta }_{\text{2}}} \right)\left( \frac{{{G}^{*}}\Lambda }{{{G}^{*}}+\lambda _{z}^{-1}{{\Lambda }^{2}}}-3\sqrt{{{\lambda }_{z}}{{G}^{*}}}\hskip 2pt{{\tan }^{-1}}\frac{\Lambda }{\sqrt{{{\lambda }_{z}}{{G}^{*}}}} \right), \notag \\ 
\Delta P_{2}^{*}\left( \Lambda  \right) & =\frac{{{\left( {{Q}^{*}} \right)}^{2}}}{2{{\lambda }_{z}}{{\left( 1+\beta _{3}^{2}{{\lambda }_{z}}{{G}^{*}} \right)}^{2}}}\left[ \beta _{3}^{2}\ln \frac{{{\lambda }_{z}}{{G}^{*}}+{{\Lambda }^{2}}}{{{\left( 1+{{\beta }_{3}}\Lambda  \right)}^{2}}} \right. \notag\\ 
& \left. +\frac{{{\beta }_{3}}\left( \beta _{3}^{2}{{\lambda }_{z}}{{G}^{*}}-1 \right)}{\sqrt{{{\lambda }_{z}}{{G}^{*}}}}\hskip 2pt{{\tan }^{-1}}\frac{\Lambda }{\sqrt{{{\lambda }_{z}}{{G}^{*}}}}+\left( 1+{{G}^{*}}\beta _{3}^{2}{{\lambda }_{z}} \right)\frac{{{\beta }_{3}}\Lambda -1}{{{\lambda }_{z}}{{G}^{*}}+{{\Lambda }^{2}}} \right], \notag\\ 
N_{r1}^{*}\left( \Lambda  \right) & =\left( {{\lambda }_{z}}+\frac{{{\mu }_{20}}}{{{\mu }_{10}}\lambda _{z}^{3}} \right){{\Lambda }^{2}}-\left( \frac{1}{{{\lambda }_{z}}}+\frac{{{\mu }_{20}}}{{{\mu }_{10}}}{{\lambda }_{z}} \right)\left( \frac{{{\Lambda }^{2}}}{{{\lambda }_{z}}}-{{G}^{*}}\ln {{\lambda }_{\theta }} \right) \notag\\ 
& +\frac{2{{\Lambda }^{3}}}{3}\left( {{\lambda }_{z}}{{\beta }_{1}}+\frac{{{\mu }_{20}}{{\beta }_{2}}}{{{\mu }_{10}}\lambda _{z}^{3}} \right)-\left( \frac{{{\beta }_{\text{1}}}}{{{\lambda }_{z}}}+\frac{{{\mu }_{20}}{{\lambda }_{z}}{{\beta }_{\text{2}}}}{{{\mu }_{10}}} \right)\left( \frac{2{{\Lambda }^{3}}}{3{{\lambda }_{z}}}+{{G}^{*}}\sqrt{{{\lambda }_{z}}{{G}^{*}}}\hskip 2pt{{\tan }^{-1}}\frac{\Lambda }{\sqrt{{{\lambda }_{z}}{{G}^{*}}}} \right), \notag\\ 
N_{r2}^{*}\left( \Lambda  \right) & =\frac{{{\left( {{Q}^{*}} \right)}^{2}}}{2\lambda _{z}^{2}\left( 1+\beta _{3}^{2}{{\lambda }_{z}}{{G}^{*}} \right)}\left[ \ln \frac{{{\lambda }_{z}}{{G}^{*}}+{{\Lambda }^{2}}}{{{\left( 1+{{\beta }_{3}}\Lambda  \right)}^{2}}}+2{{\beta }_{3}}\sqrt{{{\lambda }_{z}}{{G}^{*}}}\hskip 2pt{{\tan }^{-1}}\frac{\Lambda }{\sqrt{{{\lambda }_{z}}{{G}^{*}}}} \right].
\end{align}

Using Mathematica (Wolfram Research, Inc., 2013), we   validated the previous formulations derived from the analytical integrations. When there are no functional gradients (i.e., $\beta_i=0$) and the SEA tube is characterized by the \emph{neo-Hookean} ideal dielectric model (i.e., $\mu_{20}=0$), Eqs.~\eqref{PandNr} and \eqref{diffparts} reduce to
\begin{align}
& \Delta P=-\frac{{{\mu }_{10}}}{{{\lambda }_{z}}}\left( \ln \frac{{{\lambda }_{a}}}{{{\lambda }_{b}}}+\frac{1}{2{{\lambda }_{z}}}\frac{\lambda _{a}^{2}-\lambda _{b}^{2}}{\lambda _{a}^{2}\lambda _{b}^{2}} \right)+{{\left( \frac{Q(a)}{2\pi AL} \right)}^{2}}\frac{{{A}^{4}}\left( \eta^{2}-1 \right)}{2{{\varepsilon }_{a0}}\lambda _{z}^{3}{{a}^{2}}{{b}^{2}}},\notag\\
& {{N}_{r}}=\pi {{\mu }_{10}}{{A}^{2}}\left[ \left( {{\lambda }_{z}}-\lambda _{z}^{-2} \right)\left( \eta^{2}-1 \right)-\frac{\lambda _{a}^{2}{{\lambda }_{z}}-1}{\lambda _{z}^{2}}\ln \frac{{{\lambda }_{a}}}{{{\lambda }_{b}}} \right]-{{\left( \frac{Q(a)}{2\pi AL} \right)}^{2}}\frac{\pi {{A}^{2}}}{{{\varepsilon }_{a0}}\lambda _{z}^{2}}\ln \frac{b}{a},
\end{align}
which, for $P_{\text{out}}=0$, are equal to Eqs.~(59) and (60) obtained by \citet{melnikov2016finite} but expressed in a different notation.

Analogous to the derivation of the pressure difference {\color{black} in Eq.~\eqref{PandNr}$_1$}, integration of Eq.~\eqref{sigmarr} gives the radial normal stress as
\begin{equation} \label{taurr}
\tau _{rr}^{*}=-\left. \left[ \Delta P_{1}^{*}\left( \Lambda  \right)+\Delta P_{2}^{*}\left( \Lambda  \right) \right] \right|_{1}^{\Lambda }-P_\text{inn}^{*},
\end{equation}
where $\tau _{rr}^{*}={{\tau }_{rr}}/{{\mu }_{10}}$. Substituting Eqs.~\eqref{differentiation}$_{2,3}$ and \eqref{differentiation1} into Eq.~\eqref{universal}$_{1,2}$, we find the circumferential and axial normal stresses as
\begin{align} \label{tautheta}
& \tau _{\theta \theta }^{*}=\tau _{rr}^{*}+\left( 1+{{\beta }_{1}}\Lambda  \right)(\lambda _{\theta }^{2}-\lambda _{\theta }^{-2}\lambda _{z}^{-2})+\frac{{{\mu }_{20}}}{{{\mu }_{10}}}\left( 1+{{\beta }_{2}}\Lambda  \right)(\lambda _{\theta }^{-2}-\lambda _{\theta }^{2}\lambda _{z}^{2})-\frac{{{\left( D_{r}^{*} \right)}^{2}}}{1+{{\beta }_{3}}\Lambda }, \notag\\ 
& \tau _{zz}^{*}=\tau _{rr}^{*}+\left( 1+{{\beta }_{1}}\Lambda  \right)(\lambda _{z}^{2}-\lambda _{\theta }^{-2}\lambda _{z}^{-2})+\frac{{{\mu }_{20}}}{{{\mu }_{10}}}\left( 1+{{\beta }_{2}}\Lambda  \right)(\lambda _{z}^{-2}-\lambda _{\theta }^{2}\lambda _{z}^{2})-\frac{{{\left( D_{r}^{*} \right)}^{2}}}{1+{{\beta }_{3}}\Lambda },
\end{align}
where $\tau _{\theta \theta }^{*}={{\tau }_{\theta \theta }}/{{\mu }_{10}}$, $\tau _{zz }^{*}={{\tau }_{zz }}/{{\mu }_{10}}$ and $D_{r}^{*}={{D}_{r}}/\sqrt{{{\mu }_{10}}{{\varepsilon }_{a0}}}$. The Lagrange multiplier can be obtained by Eqs.~\eqref{MR1} and \eqref{iniconstitutive}$_1$ as
\begin{equation}
{{p}^{*}}=\left( 1+{{\beta }_{1}}\Lambda  \right)\lambda _{\theta }^{-2}\lambda _{z}^{-2}-\frac{{{\mu }_{20}}}{{{\mu }_{10}}}\left( 1+{{\beta }_{2}}\Lambda  \right)(\lambda _{\theta }^{-2}+\lambda _{z}^{-2})+\frac{{{\left( D_{r}^{*} \right)}^{2}}}{1+{{\beta }_{3}}\Lambda }-\tau _{rr}^{*},
\end{equation}
where ${{p}^{*}}=p/{{\mu }_{10}}$. The dimensionless form of the relation \eqref{electricDIS} is written as
\begin{equation} \label{D-Q}
{{\left( D_{r}^{*} \right)}^{2}}=\frac{{{\left( {{Q}^{*}} \right)}^{2}}}{\lambda _{z}^{2}\left( {{G}^{*}}+\lambda _{z}^{-1}{{\Lambda }^{2}} \right)}.
\end{equation}

Without the electro-mechanical coupling, the results obtained above recover those of \citet{batra2009inflation} and \citet{chen2017bifurcation} for the \emph{purely elastic functionally graded} tube. We note from Eqs.~\eqref{taurr}-\eqref{D-Q} that the initial physical
variables are radially inhomogeneous due to the application of a pressure difference or a radial electric voltage, even for homogeneous tubes. In the absence of pressure and voltage, a uniform deformation state always exists in functionally graded elastomeric tubes  subject to an axial stretch only. 
In that state, ${{\lambda }_{r}}={{\lambda }_{\theta }}=\lambda _{z}^{-1/2}$, the radial and circumferential stresses both vanish, and the only nonzero stress component is the axial normal stress ${{\tau }_{zz}}$, given by
\begin{equation}
{{\tau }_{zz}}={{\mu }_{1}}(R)\left( \lambda _{z}^{2}-\lambda _{z}^{-1} \right)+{{\mu }_{2}}(R)\left( \lambda _{z}^{-2}-{{\lambda }_{z}} \right),
\end{equation}
which varies along the radial direction of the functionally graded  tube.


\subsection{Numerical results}\label{section3.3}


{\color{black}In the following numerical calculations, the elastic material constants ${{\mu }_{10}}$ and ${{\mu }_{20}}$  take the values ${{\mu }_{10}}=1.858\times {{10}^{5}}$~Pa and ${{\mu }_{20}}=-0.1935\times {{10}^{5}}$~Pa, as obtained for rubber by \citet{batra2005treloar}; thus here, ${{\mu }_{20}}/{{\mu }_{10}}=-0.104$. 
The two elastic moduli gradient parameters ${{\beta }_{1}}$ and ${{\beta }_{2}}$ are assumed to be equal, ${{\beta }_{1}}={{\beta }_{2}}$.} Moreover, the requirement that the shear modulus $\mu(R)={{\mu }_{1}}(R)-{{\mu }_{2}}(R)$ and the dielectric permittivity $\varepsilon(R)$ of the functionally graded SEA material in the undeformed state should be positive leads to the following condition for the material gradient parameters: $\beta_i>-1/\eta \text{ } (i=1,2,3)$. 
Here the initial undeformed shape factor $\eta  =B/A$ is fixed as $\eta=2.0$, which gives $\beta_i>-0.5$.


\subsubsection{Nonlinear axisymmetric deformation}\label{section3.3.1}


First, we examine the nonlinear axisymmetric response of the functionally graded SEA tube to different biasing fields. The numerical results are calculated from Eqs.~\eqref{voltage-Q1}, \eqref{voltage*}, \eqref{PandNr}$_1$ and \eqref{diffparts}$_{1,2}$.

\begin{figure}[htbp]
	\centering	
	\includegraphics[width=0.94\textwidth]{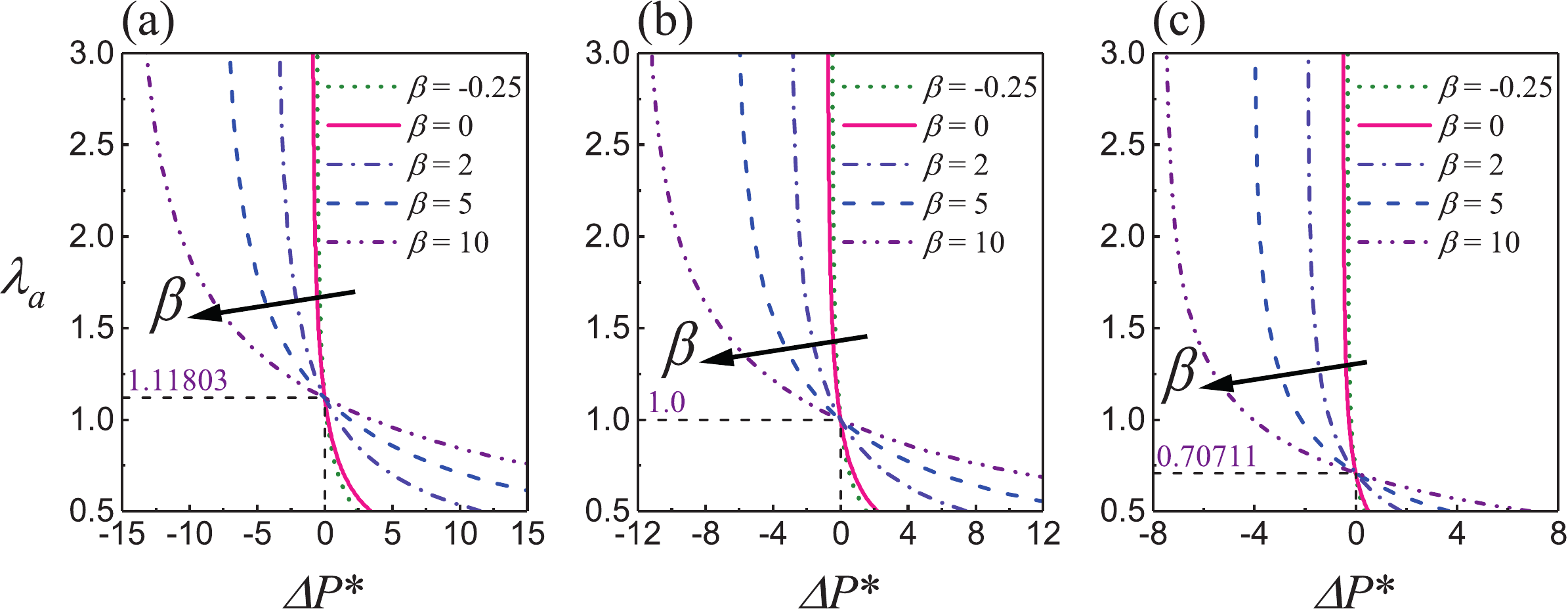}
	\caption{Variations of the inner surface circumferential ratio $\lambda_a$ with the dimensionless pressure difference $\Delta {{P}^{*}}$ in a functionally graded SEA tube with no voltage $V^*=0$, for different values of material gradient $\beta=\beta_1=\beta_2=\beta_3$ and various axial pre-stretches $\lambda_z$: (a) $\lambda_z=0.8$; (b) $\lambda_z=1.0$; (c) $\lambda_z=2.0$.}
	\label{Fig2}
\end{figure}

For three axial pre-stretches ${{\lambda }_{z}}=0.8, 1.0, 2.0$, Fig.~\ref{Fig2} shows the variations of ${{\lambda }_{a}}$ with the dimensionless pressure difference $\Delta {P}^{*}$ for various material gradient values $\beta=\beta_i \text{ } (i=1,2,3)$. 
Note that we consistently set the lower bound of ${{\lambda }_{a}}$ to be 0.5, which corresponds to a positive pressure difference. 
A smaller ${{\lambda }_{a}}$ induced by a larger positive $\Delta {P}^{*}$ may lead to an instability of the tube \citep{chen2017bifurcation}. 
As explained earlier, a uniform deformation state always exists for the functionally  graded tube when there is no pressure difference and no voltage, i.e., ${{\lambda }_{r}} = \lambda_\theta = \lambda _{z}^{-1/2} = $ constant when $\Delta {{P}^{*}}=V^*=0$. 
Specifically we find ${{\lambda }_{a}}=1.11803, 1.0, 0.70711$ when ${{\lambda }_{z}} = 0.8, 1.0, 2.0$,  respectively, as seen at the intersection point of all curves  in each of the panels of Fig.~\ref{Fig2}. 
For a fixed axial pre-stretch and fixed material gradient, ${{\lambda }_{a}}$ increases monotonically when $\Delta {P}^{*}$ decreases. 
Beyond a critical \emph{negative} pressure difference $\Delta P_{c}^{*}<0$, no solution exists for the axisymmetric deformation  and the tube collapses, as  the compressive force exceeds the mechanical resistance force of the tube. 
For example, when $\beta=5.0$, we find that the critical pressure difference is $\Delta P_{c}^{*} = -7.42, -6.20, -4.04$ for ${{\lambda }_{z}}=0.8, 1.0,  2.0$, respectively. 
This phenomenon was observed by \citet{melnikov2016finite} for a pressurized homogeneous SEA tube. What is clearly seen in Fig.~\ref{Fig2} is the gradual decline of the critical pressure difference with an increase in $\beta$ and a decrease in $\lambda_z$. 
This trend reveals that increasing the material gradient and axial compression widens the existence range of the nonlinear response. 
Moreover, in order to reach the same level of $\lambda_a$, a larger absolute value of $\Delta {{P}^{*}}$ is required when increasing the value of $\beta$, indicating that the functionally graded SEA tube is stiffened by an increasing material gradient.

\begin{figure}[htbp]
	\centering	
	\includegraphics[width=0.94\textwidth]{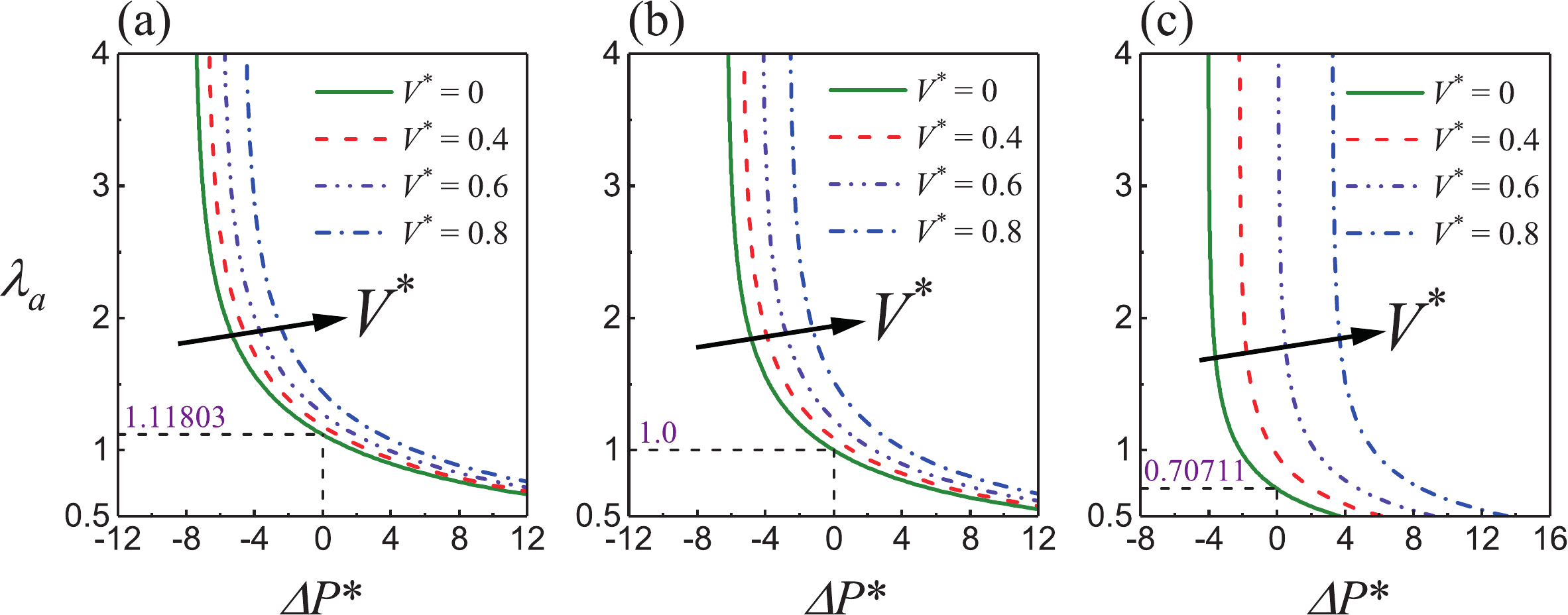}
	\caption{Variations of the inner surface circumferential ratio $\lambda_a$ with the dimensionless pressure difference $\Delta {{P}^{*}}$ in a functionally graded SEA tube with  material gradient $\beta=\beta_1=\beta_2=\beta_3=5.0$, for different values of voltage $V^*$ and various axial pre-stretches $\lambda_z$: (a) $\lambda_z=0.8$; (b) $\lambda_z=1.0$; (c) $\lambda_z=2.0$.}
	\label{Fig3}
\end{figure}

Fig.~\ref{Fig3} displays ${{\lambda }_{a}}$ as a function of $\Delta {{P}^{*}}$ for the three axial pre-stretches ${{\lambda }_{z}}=0.8, 1.0, 2.0$ and various values of increasing voltage ($V^*=0.0, 0.4, 0.6, 0.8$). 
We observe that $\lambda_a$ rises notably when the voltage increases for a fixed $\Delta {{P}^{*}}$, which indicates that the voltage tends to inflate the SEA tube. Furthermore, the critical pressure difference $\Delta P_{c}^{*}$ is lifted up with an increase in $V^*$. 
In particular, for $\lambda_z=2.0$ and a high voltage ($V^*=0.8$), $\Delta P_{c}^{*}$ becomes \emph{positive}, which means that an external pressure larger than the internal pressure is needed to maintain the axisymmetric deformation.

\begin{figure}[htbp]
	\centering	
	\includegraphics[width=0.94\textwidth]{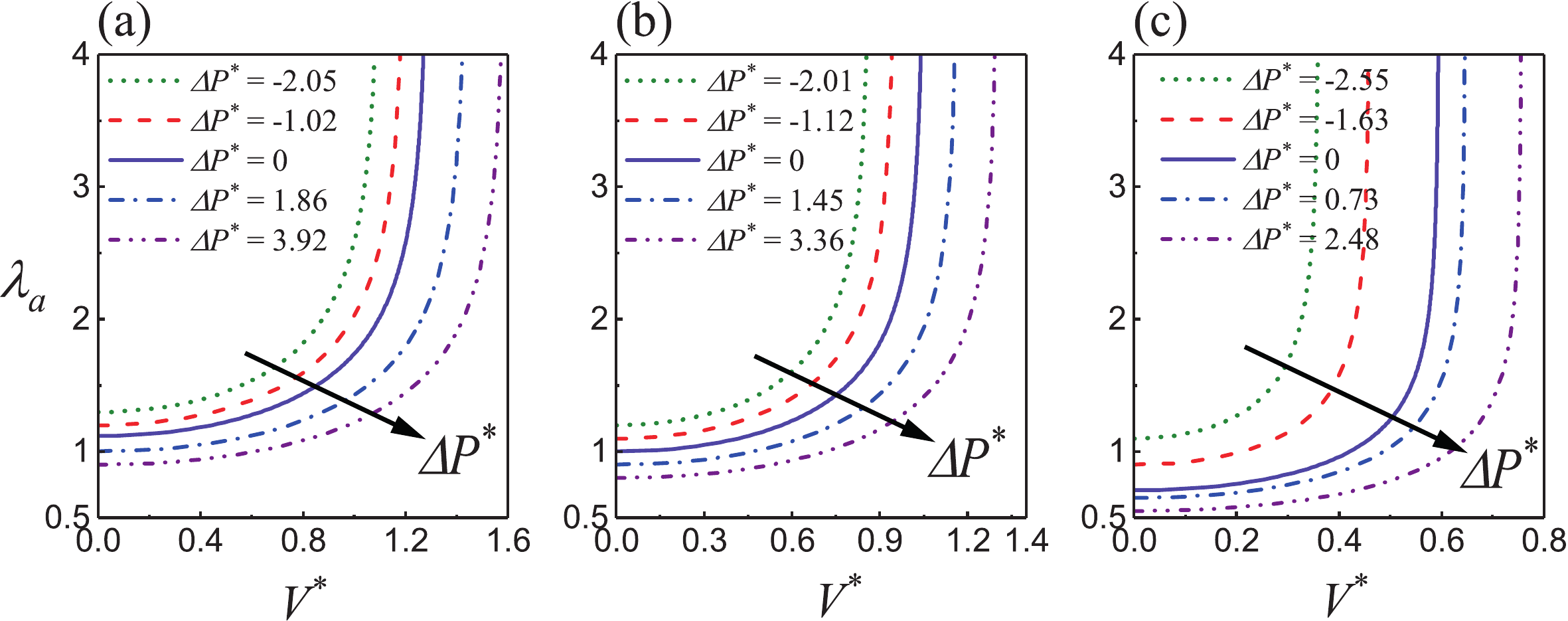}
	\caption{Variations of the inner surface circumferential ratio $\lambda_a$ with the dimensionless voltage $V^*$ in a functionally graded SEA tube with material gradient $\beta=\beta_1=\beta_2=\beta_3=5.0$, for different values of pressure difference $\Delta {{P}^{*}}$ and various axial pre-stretches $\lambda_z$: (a) $\lambda_z=0.8$; (b) $\lambda_z=1.0$; (c) $\lambda_z=2.0$.}
	\label{Fig4}
\end{figure}

For the three axial pre-stretches ${{\lambda }_{z}} = 0.8, 1.0, 2.0$, Fig.~\ref{Fig4} illustrates the variations of the inner side circumferential stretch ${{\lambda }_{a}}$ with the dimensionless voltage $V^*$, for various values of pressure difference $\Delta {{P}^{*}}$. 
It shows that there is a nonlinear monotonous increase in ${{\lambda }_{a}}$ with $V^*$, thus inflating the SEA tube. Similar to the case of pressure difference in Figs.~\ref{Fig2} and \ref{Fig3}, there is no solution of the axisymmetric deformation beyond a critical voltage $V_c^*$, once the electrostatic compressive force surpasses the elastic resistance force of the tube itself \citep{shmuel2013axisymmetric, wu2017guided}. The critical voltage value $V_c^*$ increases with the pressure difference $\Delta {{P}^{*}}$ but decreases with the axial pre-stretch $\lambda_z$. Thus, when the applied voltage becomes large, the axisymmetric deformation may be maintained by increasing the external pressure.

\begin{figure}[htbp]
	\centering	
	\includegraphics[width=0.94\textwidth]{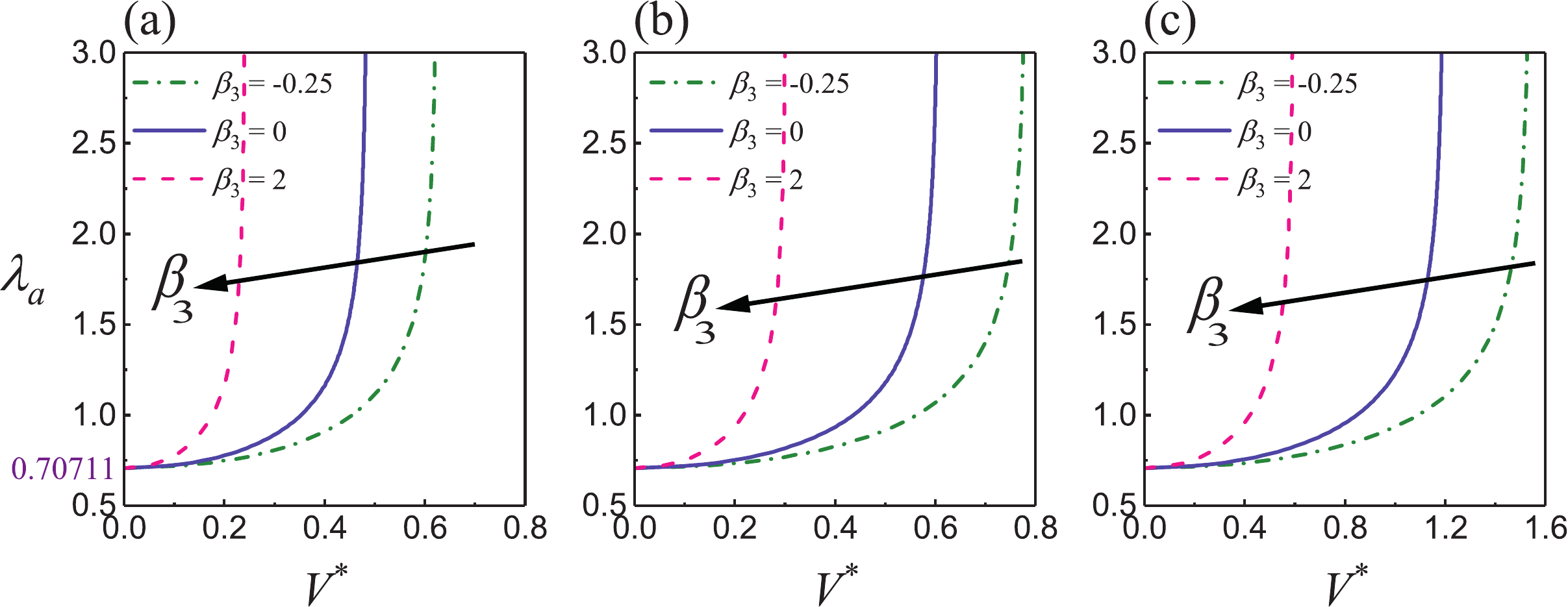}
	\caption{Variations of the inner surface circumferential ratio $\lambda_a$ with the dimensionless voltage $V^*$ in a functionally graded SEA tube with $\lambda_z=2.0$ and $\Delta {{P}^{*}}=0$, for different {permittivity} gradients $\beta_3$ and three sets of {elastic moduli} gradients: (a) $\beta_1=\beta_2=-0.25$; (b) $\beta_1=\beta_2=0$; (c) $\beta_1=\beta_2=2.0$.}
	\label{Fig5}
\end{figure}

Fig.~\ref{Fig5} highlights the influence of material gradient parameters on the nonlinear response of $\lambda_a$ with $V^*$ for a pre-stretched functionally graded SEA tube with $\lambda_z=2.0$ and $\Delta {{P}^{*}}=0$. What is striking here is the remarkable decrease in the critical voltage $V_c^*$ when  the permittivity gradient $\beta_3$ increases. 
The reason for this behaviour is that the larger the permittivity gradient, the stronger the electrostatic compressive force induced by the electric field in the tube, leading to a lower critical voltage. 
In order to reach the same level of circumferential stretch $\lambda_a$, a lower voltage may be applied to a functionally graded SEA tube with a larger permittivity gradient. Moreover, we find that for a stiffer functionally graded tube (with a larger elastic moduli gradient $\beta_1=\beta_2$), the critical voltage $V_c^*$ increases markedly. As a result, \emph{a greater actuation at a low voltage may be achieved by increasing the permittivity gradient or by decreasing the elastic moduli gradients}.

\begin{figure}[h!]
	\centering	
	\includegraphics[width=0.93\textwidth]{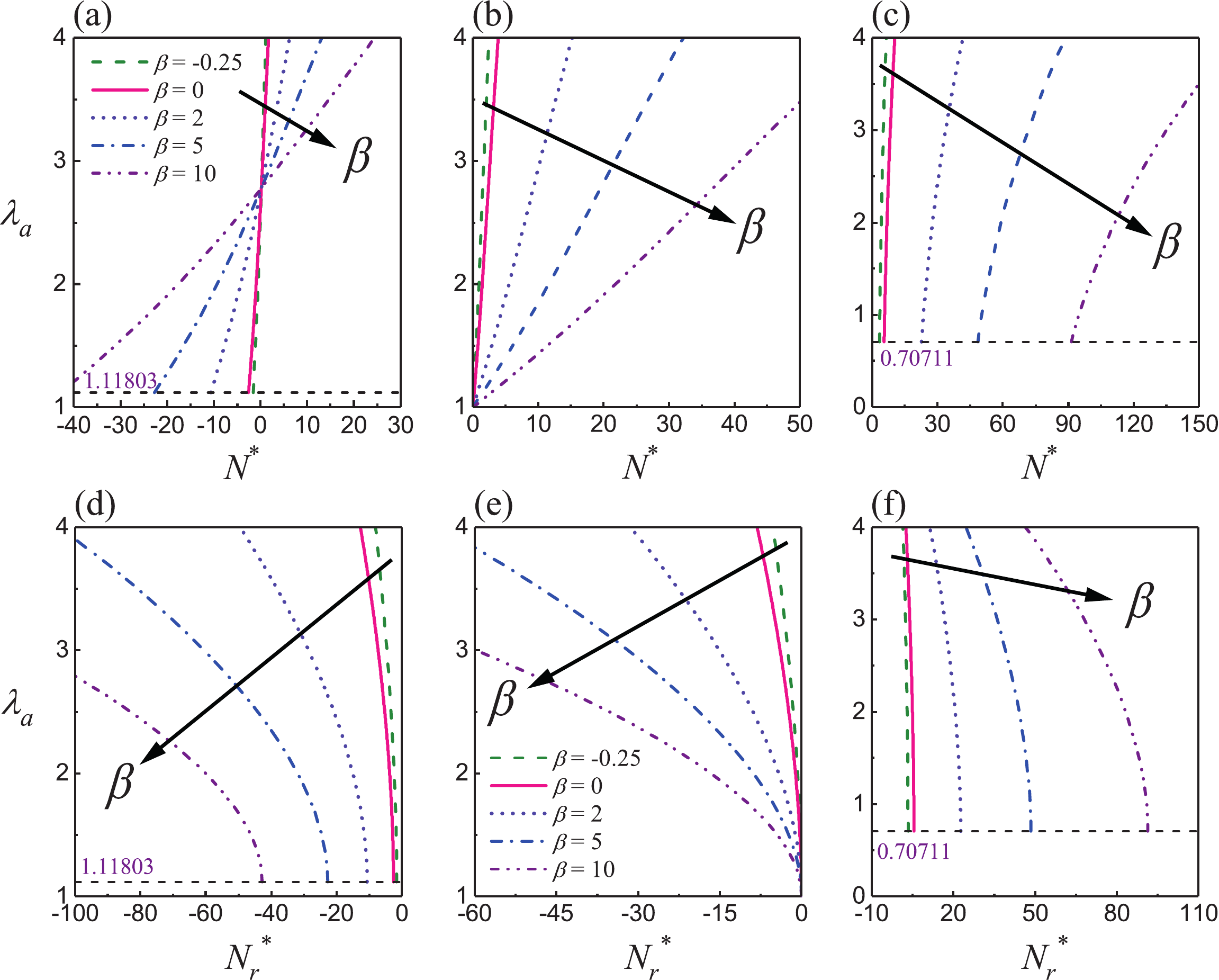}
	\caption{Variations of the inner surface circumferential ratio $\lambda_a$ with the dimensionless \emph{resultant} axial force $N^{*}$ (a-c) and \emph{reduced} axial force $N_{r}^{*}$ (d-f) in a functionally graded SEA tube with ${{P}_{\text{out}}^{*}}=0$ and $V^*=0$, for different material gradients $\beta=\beta_1=\beta_2=\beta_3$ and various axial pre-stretches $\lambda_z$: (a, d) $\lambda_z=0.8$; (b, e) $\lambda_z=1.0$; (c, f) $\lambda_z=2.0$.}
	\label{Fig6}
\end{figure}

We emphasize that according to Eqs.~\eqref{voltage-Q1}, \eqref{voltage*}, \eqref{PandNr} and \eqref{diffparts}, the nonlinear axisymmetric deformation and the reduced axial force $N_{r}$ are independent of the specific value of the internal or external pressure, and depend only on the material gradients, axial pre-stretch, voltage, and pressure difference. 
The actual internal or external pressure, however,  affects the \emph{resultant} axial force $N$ and the initial stresses ${{\tau }_{ii}}\text{ }(i=r,\theta ,z)$, see Eqs.~\eqref{PandNr}$_2$, \eqref{taurr} and \eqref{tautheta}. The effect of the circumferential stretch ${{\lambda }_{a}}$ on the dimensionless \emph{resultant} ($N^*$) and \emph{reduced} ($N_{r}^*$) axial forces in a functionally graded SEA tube is illustrated in Fig.~\ref{Fig6} for the same material gradients $\beta$ and axial pre-stretches ${{\lambda }_{z}}$ as those in Fig.~\ref{Fig2}. 
Note that there is no external pressure here (i.e., ${{P}_{\text{out}}}=0$) and only an internal pressure ${{P}_\text{inn}}$ is applied on the inner surface of the tube. Here the increase of ${{\lambda }_{a}}$ reflects the inflation of the tube resulting from the increase in ${{P}_\text{inn}}$. 
The starting point of ${{\lambda }_{a}}$ corresponding to ${{P}_{\text{inn}}}=V^*=0$ satisfies the relation ${{\lambda }_{a}}=\lambda _{z}^{-1/2}$.

In the case where the tube has \emph{open ends}, a resultant axial force $N^*$ needs to be applied to maintain a fixed axial pre-stretch $\lambda_z$, as shown in Figs.~\ref{Fig6}(a)-(c). 
For an axial contraction, ${{\lambda }_{z}}=0.8$ in Fig.~\ref{Fig6}(a), a negative (compressive) resultant axial force  is required. 
As the inflation takes place, the axial force eventually becomes positive (tensile) once a sufficient value of ${{\lambda }_{a}}$ is reached. 
In contrast, a positive (tensile) axial force is initially required for an axial extension  ${{\lambda }_{z}}=2.0$, which then increases continuously with the inflation, as displayed in Fig.~\ref{Fig6}(c). 
For ${{\lambda }_{z}}=1.0$ in Fig.~\ref{Fig6}(b), there is no axial force initially, but an increasing tensile axial force is then required during the inflation. 
We also observe from Figs.~\ref{Fig6}(a)-(c) that an increase in the material gradient significantly increases the  absolute value of ${{N}^{*}}$ required to reach the same level of ${{\lambda }_{a}}$, due to the material stiffening effect.

When the tube has \emph{closed ends}, the externally applied axial load is defined as the reduced axial force $N_r$ in Eq.~\eqref{axial-forceRed}. 
We present the variation of $N_r^*$ with $\lambda_a$ (i.e., ${{P}_\text{inn}}$) in Figs.~\ref{Fig6}(d)-(f). 
We observe that a stiffer tube with a larger material gradient $\beta$ requires a larger absolute value of ${{N}_r^{*}}$ to reach the same level of ${{\lambda }_{a}}$, a  phenomenon similar to that observed in Figs.~\ref{Fig6}(a)-(c). 
However, the reduced axial force $N_r^*$ \emph{decreases monotonically} with $\lambda_a$, which is in contrast to the behaviour of  the resultant force $N^*$. This is because the increase of internal pressure accompanied by an increase in $\lambda_a$ makes the tube with closed ends exhibit both inflation and elongation trends, and the decreasing axial force caused by the elongation trend prevails over the increasing inflation-induced axial force (also see Eq.~\eqref{axial-forceRed}).


\subsubsection{Inhomogeneous biasing fields}\label{section3.3.2}


Next, we examine the effect of the applied voltage and of the material gradients on the inhomogeneous biasing fields. 

\begin{figure}[h!]
	\centering	
	\includegraphics[width=0.92\textwidth]{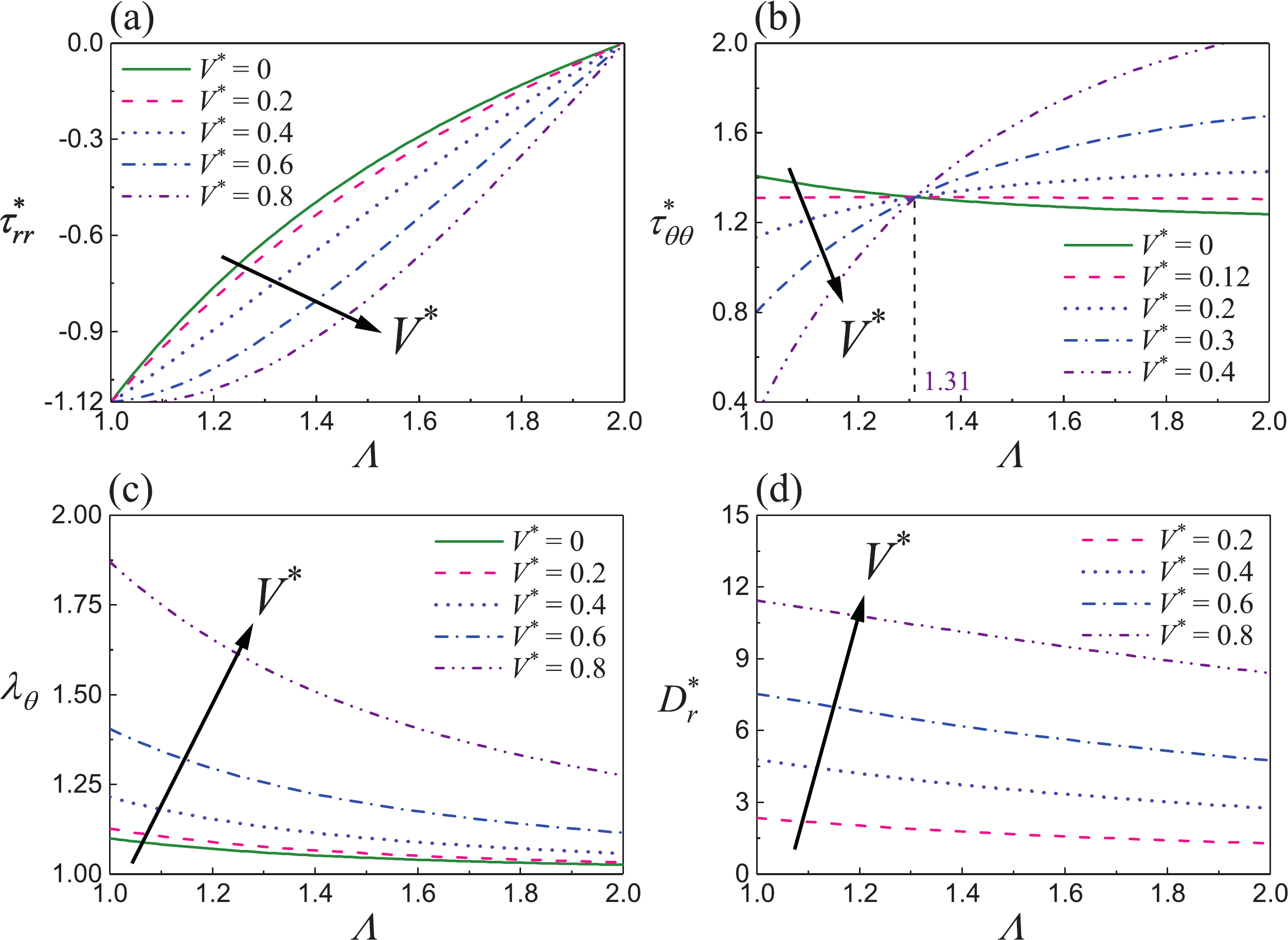}
	\caption{Radial distributions of the dimensionless radial normal stress $\tau_{rr}^*$ (a), circumferential normal stress $\tau_{\theta \theta}^*$ (b), circumferential stretch $\lambda_\theta$ (c), and radial electric displacement $D_r^*$ (d) in a functionally graded SEA tube with ${{P}_{\text{out}}^{*}}=0$, ${{P}_{\text{inn}}^{*}}=1.12$, $\lambda_z=1.0$ and $\beta_1=\beta_2=\beta_3=5.0$, for different values of $V^*$.}
	\label{Fig7}
\end{figure}

Fig.~\ref{Fig7} is plotted according to Eqs.~\eqref{lamdaab}, \eqref{taurr}, \eqref{tautheta}$_1$ and \eqref{D-Q}.
It illustrates the variations of $\tau_{rr}^*$, $\tau_{\theta \theta}^*$, $\lambda_\theta$ and $D_r^*$ along the radial direction ($\Lambda =R/A$) in a functionally graded SEA tube subject to the internal pressure ${{P}_\text{inn}^*}=1.12$, for different values of voltage $V^*$.
It is clear that these biasing fields are radially inhomogeneous in the tube, even for $V^*=0$, where $D_r^*$ disappears. 
As expected, the radial stress component $\tau_{rr}^*$  is compressive throughout the tube and has its minimum value at the inner surface, independent of the applied voltage. 
The circumferential stress component  $\tau_{\theta \theta}^*$, on the other hand, is tensile in the tube and its radial distribution is affected by the voltage. 
Specifically, the maximum value of $\tau_{\theta \theta}^*$ appears at the inner surface for a low voltage and at the outer surface for a higher voltage, as expected from Eq.~\eqref{tautheta}$_1$. 
The radial distribution of $\tau_{\theta \theta}^*$ is almost uniform for a moderate voltage, as seen for $V^*=0.12$ in Fig.~\ref{Fig7}(b). Therefore, as the voltage increases, the degree of inhomogeneity of $\tau_{\theta \theta}^*$ decreases first and then increases. Another interesting phenomenon is that $\tau_{\theta \theta}^*$ remains essentially unchanged with the applied voltage at the point $\Lambda \simeq 1.31$. Figs.~\ref{Fig7}(c)-(d) show that increasing the electric voltage enlarges substantially the inhomogeneous degree of $\lambda_\theta$ and $D_r^*$, and that their values at the inner surface are always larger than those at the outer surface, which is  in agreement with Eqs.~\eqref{lamdaab} and \eqref{D-Q}.

\begin{figure}[h!]
	\centering	
	\includegraphics[width=0.95\textwidth]{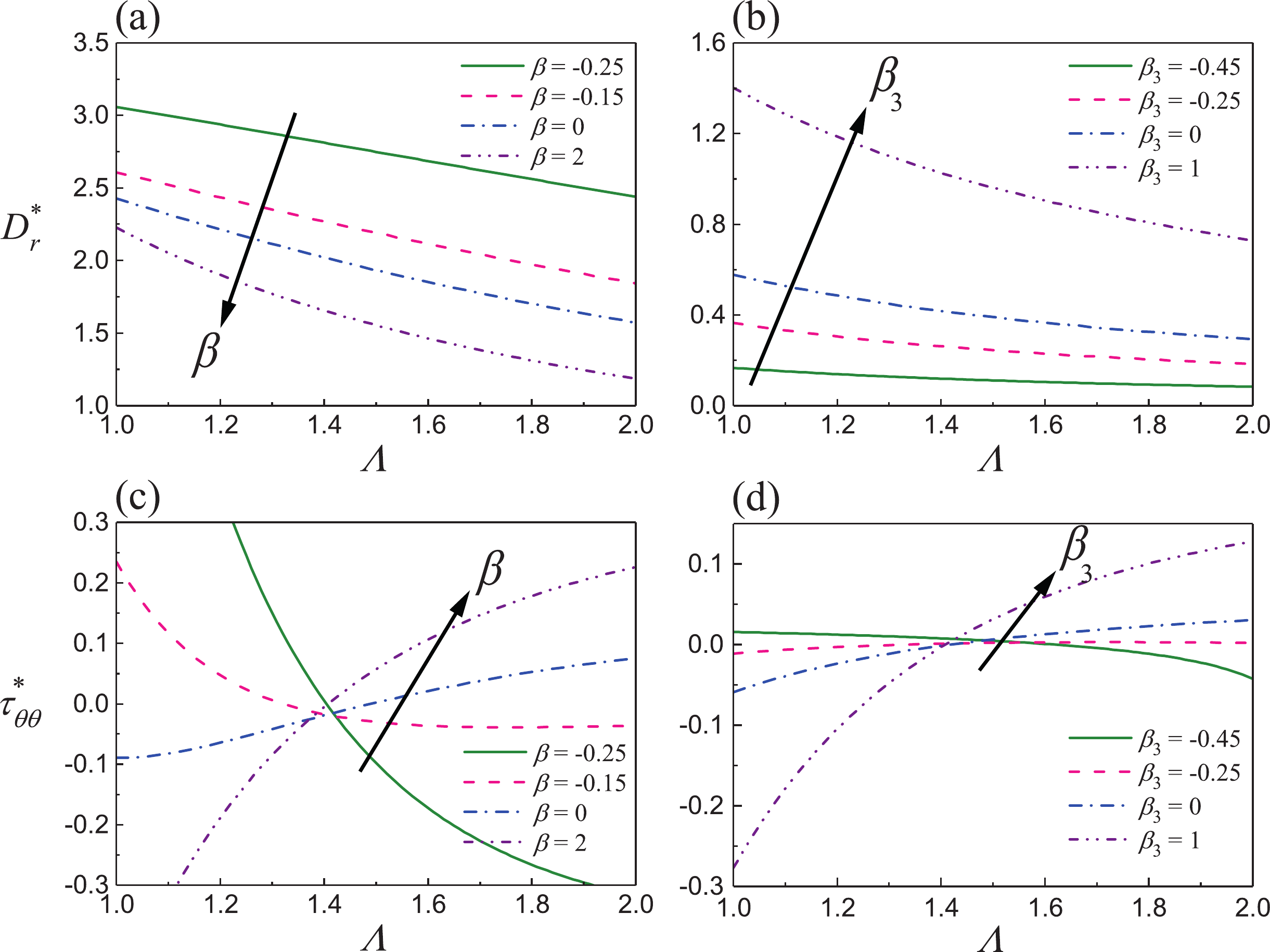}
	\caption{Radial variations of the dimensionless electric displacement $D_r^*$ (a, b) and circumferential normal stress $\tau_{\theta \theta}^*$ (c, d) in a functionally graded SEA tube with ${{P}_{\text{out}}^{*}}={{P}_{\text{inn}}^{*}}=0$, $\lambda_z=1.0$ and $V^*=0.4$: (a, c) for  fixed  {permittivity} gradient $\beta_3=2.0$ and different {elastic moduli} gradients $\beta=\beta_1=\beta_2$; (b, d) for fixed {elastic moduli} gradients $\beta_1=\beta_2=2.0$ and different {permittivity} gradients $\beta_3$.}
	\label{Fig8}
\end{figure}

Fig.~\ref{Fig8} shows the effect of the elastic moduli gradients $\beta=\beta_1=\beta_2$ and the permittivity gradient $\beta_3$ on the radial distributions of $\tau_{\theta \theta}^*$ and $D_r^*$ in a functionally graded SEA tube subject to the voltage $V^*=0.4$. We see from Figs.~\ref{Fig8}(a)-(b) that the value of $D_r^*$ drops down with an increase in the elastic moduli gradient $\beta=\beta_1=\beta_2$ and goes up with an increasing permittivity gradient $\beta_3$. 
This is because the value of $D_r^*$  depends mainly on the denominator in the fraction of Eq.~\eqref{voltage-Q1}, where a larger $\beta$ leads to a smaller $\lambda_a$ for a given voltage (i.e., a smaller ${{G}^{*}}=\lambda _{a}^{2}-\lambda _{z}^{-1}$) while a larger $\beta_3$ results in the opposite trend (also seen in Fig.~\ref{Fig5}). 
Figs.~\ref{Fig8}(c)-(d) clearly show that the material gradients also influence noticeably the distribution of $\tau_{\theta \theta}^*$. Similar to the role of voltage in Fig.~\ref{Fig7}(b), as $\beta$ or $\beta_3$ increases, the inhomogeneous degree of $\tau_{\theta \theta}^*$ decreases first and then becomes large. 
In particular, the radial distribution of $\tau_{\theta \theta}^*$ when $\beta_1=\beta_2=2.0$ and $\beta_3 =-0.25$ is nearly uniform. Thus, \emph{the stress concentration or sharp stress variations in SEA tubes subject to electric fields may be alleviated by properly tailoring the material gradient parameters}. Interestingly, the circumferential stress $\tau_{\theta \theta}^*$ takes almost the same value at the point $\Lambda =\sqrt{\eta }\simeq1.414$ when varying $\beta$ or $\beta_3$, which has been observed by \citet{batra2009inflation} for purely elastic functionally graded tubes.

\section{State-space method for incremental fields} \label{sec4}


Because of the functional gradients and of the radially inhomogeneous character of the biasing fields for the static nonlinear deformation in the SEA tube shown in Sec.~\ref{section3}, the instantaneous electro-elastic moduli also depend on the radial coordinate $r$. Thus, it is intractable to solve analytically the resulting incremental governing equations by means of the conventional displacement-based method because they are a system of coupled partial differential equations with \emph{variable coefficients}. 
Moreover, their numerical resolution may run into problems because they are likely to be stiff.
{\color{black}Note that \citet{shmuel2013axisymmetric} used the compound matrix method to study  axisymmetric waves propagating in neo-Hookean ideal SEA tubes under  radial electric voltage, but  met the problem of numerical divergence when searching for roots in the case of thick-walled tubes.}

In this work, we rely on the \textit{state-space method} (SSM) to combine the incremental state-space formalism (Sec.~\ref{section4.1}) with the approximate laminate technique (Sec.~\ref{section4.2}) in order to derive the dispersion relations of the superimposed axisymmetric waves propagating in the deformed functionally graded SEA tube. {\color{black}The SSM, as a special mixed-variable method, transforms the governing equations into a set of first-order ordinary differential equations with respect to one particular coordinate, the radial coordinate here.} The SSM presents several advantages over the displacement-based methods in solving many practical problems, such as multi-layered or functionally graded structures \citep{chen2017bifurcation, wu2018propagation} and  inhomogeneous biasing fields \citep{wu2017guided, mao2019electrostatically}, as it is numerically robust. {\color{black}The interested readers are referred to \citet{ding2001three} for more details and the references cited therein.}


\subsection{Incremental equations and state-space formalism in cylindrical coordinates}
\label{section4.1}


In the cylindrical coordinates $\left( r,\theta ,z \right)$ shown in Fig.~\ref{Fig1}(b), the basic incremental governing equations of the deformed tube are the incremental equations of motion and incremental Gauss's law,	
	\begin{align} \label{incmotion}
	& \frac{\partial {{{\dot{T}}}_{0rr}}}{\partial r}+\frac{1}{r}\frac{\partial {{{\dot{T}}}_{0\theta r}}}{\partial \theta }+\frac{{{{\dot{T}}}_{0rr}}-{{{\dot{T}}}_{0\theta \theta }}}{r}+\frac{\partial {{{\dot{T}}}_{0zr}}}{\partial z}=\rho \frac{{{\partial }^{2}}{{u}_{r}}}{\partial {{t}^{2}}}, \notag \\ 
	& \frac{\partial {{{\dot{T}}}_{0r\theta }}}{\partial r}+\frac{1}{r}\frac{\partial {{{\dot{T}}}_{0\theta \theta }}}{\partial \theta }+\frac{{{{\dot{T}}}_{0\theta r}}+{{{\dot{T}}}_{0r\theta }}}{r}+\frac{\partial {{{\dot{T}}}_{0z\theta }}}{\partial z}=\rho \frac{{{\partial }^{2}}{{u}_{\theta }}}{\partial {{t}^{2}}}, \notag \\ 
	& \frac{\partial {{{\dot{T}}}_{0rz}}}{\partial r}+\frac{1}{r}\frac{\partial {{{\dot{T}}}_{0\theta z}}}{\partial \theta }+\frac{\partial {{{\dot{T}}}_{0zz}}}{\partial z}+\frac{{{{\dot{T}}}_{0rz}}}{r}=\rho \frac{{{\partial }^{2}}{{u}_{z}}}{\partial {{t}^{2}}}, \notag \\
	&\frac{\partial {{{\dot{\mathcal{D}}}}_{0r}}}{\partial r}+\frac{1}{r}\left( \frac{\partial {{{\dot{\mathcal{D}}}}_{0\theta }}}{\partial \theta }+{{{\dot{\mathcal{D}}}}_{0r}} \right)+\frac{\partial {{{\dot{\mathcal{D}}}}_{0z}}}{\partial z}=0,
	\end{align}
together with the incremental incompressibility constraint,
	\begin{equation} \label{incincom}
	\frac{\partial {{u}_{r}}}{\partial r}+\frac{1}{r}\left( \frac{\partial {{u}_{\theta }}}{\partial \theta }+{{u}_{r}} \right)+\frac{\partial {{u}_{z}}}{\partial z}=0,
	\end{equation}
and the incremental constitutive relations,
	\begin{align} 
& {{{\dot{T}}}_{0rr}}={{c}_{11}}\frac{\partial {{u}_{r}}}{\partial r}+{{c}_{12}}\frac{1}{r}\left( \frac{\partial {{u}_{\theta }}}{\partial \theta }+{{u}_{r}} \right)+{{c}_{13}}\frac{\partial {{u}_{z}}}{\partial z}+{{e}_{11}}\frac{\partial \dot{\phi }}{\partial r}-\dot{p}, \notag\\ 
& {{{\dot{T}}}_{0\theta \theta }}={{c}_{12}}\frac{\partial {{u}_{r}}}{\partial r}+{{c}_{22}}\frac{1}{r}\left( \frac{\partial {{u}_{\theta }}}{\partial \theta }+{{u}_{r}} \right)+{{c}_{23}}\frac{\partial {{u}_{z}}}{\partial z}+{{e}_{12}}\frac{\partial \dot{\phi }}{\partial r}-\dot{p}, \notag\\ 
& {{{\dot{T}}}_{0zz}}={{c}_{13}}\frac{\partial {{u}_{r}}}{\partial r}+{{c}_{23}}\frac{1}{r}\left( \frac{\partial {{u}_{\theta }}}{\partial \theta }+{{u}_{r}} \right)+{{c}_{33}}\frac{\partial {{u}_{z}}}{\partial z}+{{e}_{13}}\frac{\partial \dot{\phi }}{\partial r}-\dot{p}, \notag\\ 
& {{{\dot{T}}}_{0\theta z}}={{c}_{44}}\frac{1}{r}\frac{\partial {{u}_{z}}}{\partial \theta }+{{c}_{47}}\frac{\partial {{u}_{\theta }}}{\partial z},\qquad {{{\dot{T}}}_{0z\theta }}={{c}_{47}}\frac{1}{r}\frac{\partial {{u}_{z}}}{\partial \theta }+{{c}_{77}}\frac{\partial {{u}_{\theta }}}{\partial z}, \notag\\ 
& {{{\dot{T}}}_{0rz}}={{c}_{55}}\frac{\partial {{u}_{z}}}{\partial r}+{{c}_{58}}\frac{\partial {{u}_{r}}}{\partial z}+{{e}_{35}}\frac{\partial \dot{\phi }}{\partial z},\qquad {{{\dot{T}}}_{0zr}}={{c}_{58}}\frac{\partial {{u}_{z}}}{\partial r}+{{c}_{88}}\frac{\partial {{u}_{r}}}{\partial z}+{{e}_{35}}\frac{\partial \dot{\phi }}{\partial z}, \notag\\ 
& {{{\dot{T}}}_{0r\theta }}={{c}_{66}}\frac{\partial {{u}_{\theta }}}{\partial r}+{{c}_{69}}\frac{1}{r}\left( \frac{\partial {{u}_{r}}}{\partial \theta }-{{u}_{\theta }} \right)+{{e}_{26}}\frac{1}{r}\frac{\partial \dot{\phi }}{\partial \theta }, \notag\\ 
& {{{\dot{T}}}_{0\theta r}}={{c}_{69}}\frac{\partial {{u}_{\theta }}}{\partial r}+{{c}_{99}}\frac{1}{r}\left( \frac{\partial {{u}_{r}}}{\partial \theta }-{{u}_{\theta }} \right)+{{e}_{26}}\frac{1}{r}\frac{\partial \dot{\phi }}{\partial \theta },
	\notag \\
	& {{{\dot{\mathcal{D}}}}_{0r}}={{e}_{11}} \frac{\partial {{u}_{r}}}{\partial r}+{{e}_{12}}\frac{1}{r}\left( \frac{\partial {{u}_{\theta }}}{\partial \theta }+{{u}_{r}} \right)+{{e}_{13}}\frac{\partial {{u}_{z}}}{\partial z}-{{\varepsilon }_{11}}\frac{\partial \dot{\phi }}{\partial r}, \notag\\ 
	& {{{\dot{\mathcal{D}}}}_{0\theta }}={{e}_{26}}\left[ \frac{1}{r}\left( \frac{\partial {{u}_{r}}}{\partial \theta }-{{u}_{\theta }} \right)+\frac{\partial {{u}_{\theta }}}{\partial r} \right]-{{\varepsilon }_{22}}\frac{1}{r}\frac{\partial \dot{\phi }}{\partial \theta }, \notag \\ 
	& {{{\dot{\mathcal{D}}}}_{0z}}={{e}_{35}}\left( \frac{\partial {{u}_{z}}}{\partial r}+\frac{\partial {{u}_{r}}}{\partial z} \right)-{{\varepsilon }_{33}}\frac{\partial \dot{\phi }}{\partial z}.  \label{incconsti2}
	\end{align}

Note that ${{\dot{T}}}_{0ij}\neq{{\dot{T}}}_{0ji} \,(i\neq j; i,j=r,\theta,z)$.
Also, $c_{ij}$, $e_{ij}$ and $\varepsilon_{ij}$ in Eq.~\eqref{incconsti2} are the effective material parameters associated with the instantaneous electro-elastic moduli ${{\mathcal{A}}_{0ijkl}}$, ${{\mathcal{M}}_{0ijk}}$ and ${{\mathcal{R}}_{0ij}}$ and their expressions are given in Eq.~(41) in the paper of \citet{wu2017guided}. In Eq.~\eqref{incconsti2}, we  specialized the incremental displacement gradient tensor $\mathbf{H}$ to the cylindrical coordinates. 
We also introduced an \emph{incremental electric potential} $\dot{\phi }$, defined by ${{\bm{\dot{\mathcal{E}}}}_{0}}=-\text{grad}\dot{\phi }$, so that the incremental Faraday's law {\color{black}  Eq.~\eqref{incre-governEQ}$_3$} is  identically satisfied.

For the nonlinear axisymmetric deformation of the homogeneous SEA tube subject to a radial electric displacement, \citet{wu2017guided}  derived the nonzero components of the instantaneous electro-elastic moduli tensors ${{\mathcal{A}}_{0}}$, ${{\mathcal{M}}_{0}}$ and ${{\mathcal{R}}_{0}}$ (see their Appendix B for specific expressions). 
We emphasize that tailoring the material gradient and tuning the electro-mechanical biasing fields  alters greatly the instantaneous material properties of the tube, with notable knock-on effects on the dynamic behavior of the incremental motions, as we show below.

First we write the basic incremental governing equations \eqref{incmotion}-\eqref{incconsti2} as a first-order system of  differential equations with respect to the deformed radial coordinate, in the form
\begin{equation} \label{stateeq}
\frac{\partial \mathbf{Y}}{\partial r}=\mathbf{MY},
\end{equation}
which is called the state equation, where the incremental state vector $\mathbf{Y}$ is defined as
\begin{equation} \label{statvec}
\mathbf{Y}=\left[ \begin{matrix}
{{u}_{r}} & {{u}_{\theta }} & {{u}_{z}} & {\dot{\phi }} & {{{\dot{T}}}_{0rr}} & {{{\dot{T}}}_{0r\theta }} & {{{\dot{T}}}_{0rz}} & {{{\dot{\mathcal{D}}}}_{0r}}
\end{matrix} \right]^{\text{T}},
\end{equation}
with the elements being the state variables.
The specific expressions of the $8 \times 8$ system matrix ${\mathbf{M}}$ are the same as those for a homogeneous SEA tube presented in Appendix C in \citet{wu2017guided}; they are omitted here for brevity.

Note that the state equation \eqref{stateeq} is valid for any arbitrary energy density function of incompressible, isotropic, functionally graded SEA tubes.


\subsection{Approximate laminate technique} \label{section4.2}


For axisymmetric wave propagation along the axial direction, $\partial /\partial \theta =0$, which splits the state equation \eqref{stateeq} into two systems,
\begin{equation} \label{stateaxi}
\frac{\partial {{\mathbf{Y}}_{1}}}{\partial r}={{\mathbf{M}}_{1}}{{\mathbf{Y}}_{1}},
\qquad 
\frac{\partial {{\mathbf{Y}}_{2}}}{\partial r}={{\mathbf{M}}_{2}}{{\mathbf{Y}}_{2}},
\end{equation}
where ${{\mathbf{Y}}_{1}}={{\left[ {{u}_{\theta }},{{{\dot{T}}}_{0r\theta }} \right]}^{\text{T}}}$, ${{\mathbf{Y}}_{2}}={{\left[ {{u}_{r}},{{u}_{z}},\dot{\phi },{{{\dot{T}}}_{0rr}}, {{{\dot{T}}}_{0rz}},{{{\dot{\mathcal{D}}}}_{0r}} \right]}^{\text{T}}}$ and
\renewcommand{\arraystretch}{2}
\begin{equation} \label{system1}
{{\mathbf{M}}_{1}}=\left[ \begin{matrix}
\dfrac{{{c}_{69}}}{{{c}_{66}}}\dfrac{1}{r} & \dfrac{1}{{{c}_{66}}}  \\
\rho \dfrac{{{\partial }^{2}}}{\partial {{t}^{2}}}+\dfrac{{{q}_{7}}}{{{r}^{2}}}-{{c}_{77}}\dfrac{{{\partial }^{2}}}{\partial {{z}^{2}}} & -\left( \dfrac{{{c}_{69}}}{{{c}_{66}}}+1 \right)\dfrac{1}{r}  \\
\end{matrix} \right],
\end{equation}
\begin{equation} \label{system2}
{{\mathbf{M}}_{2}}=\left[ \begin{matrix}
-\dfrac{1}{r} & -\dfrac{\partial }{\partial z} & 0 & 0 & 0 & 0  \\
-\dfrac{{{c}_{58}}}{{{c}_{55}}}\dfrac{\partial }{\partial z} & 0 & -\dfrac{{{e}_{35}}}{{{c}_{55}}}\dfrac{\partial }{\partial z} & 0 & \dfrac{1}{{{c}_{55}}} & 0  \\
\dfrac{{{q}_{1}}}{r} & {{q}_{2}}\dfrac{\partial }{\partial z} & 0 & 0 & 0 & -\dfrac{1}{{{\varepsilon }_{11}}}  \\
\rho \dfrac{{{\partial }^{2}}}{\partial {{t}^{2}}}+\dfrac{{{q}_{3}}}{{{r}^{2}}}-{{q}_{9}}\dfrac{{{\partial }^{2}}}{\partial {{z}^{2}}} & \dfrac{{{q}_{4}}}{r}\dfrac{\partial }{\partial z} & -{{q}_{10}}\dfrac{{{\partial }^{2}}}{\partial {{z}^{2}}} & 0 & -\dfrac{{{c}_{58}}}{{{c}_{55}}}\dfrac{\partial }{\partial z} & -\dfrac{{{q}_{1}}}{r}  \\
-\dfrac{{{q}_{5}}}{r}\dfrac{\partial }{\partial z} & \rho \dfrac{{{\partial }^{2}}}{\partial {{t}^{2}}}-{{q}_{6}}\dfrac{{{\partial }^{2}}}{\partial {{z}^{2}}} & 0 & -\dfrac{\partial }{\partial z} & -\dfrac{1}{r} & {{q}_{2}}\dfrac{\partial }{\partial z}  \\
-{{q}_{10}}\dfrac{{{\partial }^{2}}}{\partial {{z}^{2}}} & 0 & {{q}_{12}}\dfrac{{{\partial }^{2}}}{\partial {{z}^{2}}} & 0 & -\dfrac{{{e}_{35}}}{{{c}_{55}}}\dfrac{\partial }{\partial z} & -\dfrac{1}{r}  \\
\end{matrix} \right],
\end{equation}
\renewcommand{\arraystretch}{1}
(the material parameters $q_i$ are given in Appendix C of \cite{wu2017guided}.)

It is clear from Eqs.~\eqref{stateaxi}-\eqref{system2} that the two state variables ${{u}_{\theta }}$ and ${{{\dot{T}}}_{0r\theta }}$ are uncoupled from the other six physical variables ${{u}_{r}}$, ${{u}_{z}}$, $\dot{\phi }$, ${{{\dot{T}}}_{0rr}}$, ${{{\dot{T}}}_{0rz}}$ and ${{{\dot{\mathcal{D}}}}_{0r}}$, which indicates that there exist two classes of independent axisymmetric waves superimposed on the underlying nonlinear pre-deformation: (i) \emph{purely torsional waves} (T waves) described by ${{\mathbf{Y}}_{1}}$ and ${{\mathbf{M}}_{1}}$, with the only mechanical displacement component ${{u}_{\theta }}$ (see Fig.~\ref{Fig1}(c)); (ii) \emph{torsionless longitudinal waves} (L waves) associated with ${{\mathbf{Y}}_{2}}$ and ${{\mathbf{M}}_{2}}$, whose nonzero displacement components are ${{u}_{r}}$ and ${{u}_{z}}$ (see Fig.~\ref{Fig1}(d)). 

For time-harmonic axisymmetric waves, the traveling wave solutions are assumed in the form
\begin{align} \label{solutions}
& {{u}_{r}}=a{{U}_{r}}({\xi })\exp [\text{i}(kz-\omega t)],&& {{u}_{\theta }}=a{{U}_{\theta }}({\xi })\exp [\text{i}(kz-\omega t)], \notag\\ 
& {{u}_{z}}=a{{U}_{z}}({\xi })\exp [\text{i}(kz-\omega t)],&& \dot{\phi }=a\sqrt{{{\mu }_{10}}/{{\varepsilon }_{a0}}}\Phi ({\xi })\exp [\text{i}(kz-\omega t)] \notag\\ 
& {{{\dot{T}}}_{0rr}}={{\mu }_{10}}{{\Sigma }_{0rr}}({\xi })\exp [\text{i}(kz-\omega t)],&& {{{\dot{T}}}_{0r\theta }}={{\mu }_{10}}{{\Sigma }_{0r\theta }}({\xi })\exp [\text{i}(kz-\omega t)], \notag\\ 
& {{{\dot{T}}}_{0rz}}={{\mu }_{10}}{{\Sigma }_{0rz}}({\xi })\exp [\text{i}(kz-\omega t)],&& {{{\dot{\mathcal{D}}}}_{0r}}=\sqrt{{{\mu }_{10}}{{\varepsilon }_{a0}}}{{\Delta }_{0r}}({\xi })\exp [\text{i}(kz-\omega t)],
\end{align}
where $k$ is the axial wave number, $\omega$ is the circular frequency, $\text{i}=\sqrt{-1}$ is the imaginary unit, and $\xi=r/a$ is the \emph{dimensionless radial coordinate} in the deformed configuration.

Substituting Eq.~\eqref{solutions} into Eqs.~\eqref{stateaxi}-\eqref{system2}, we obtain
\begin{equation} \label{dimen-state-equation}
\frac{\text{d}}{\text{d}{\xi }}{{\mathbf{V}}_{1}}({\xi })={{\mathbf{\overline{M}}}_{1}}({\xi }){{\mathbf{V}}_{1}}({\xi }),
\qquad\frac{\text{d}}{\text{d}{\xi }}{{\mathbf{V}}_{2}}({\xi })={{\mathbf{\overline{M}}}_{2}}({\xi }){{\mathbf{V}}_{2}}({\xi }),
\end{equation}
where ${{\mathbf{V}}_{1}}={{\left[ {{U}_{\theta }},{{\Sigma }_{0r\theta }} \right]}^{\text{T}}}$, ${{\mathbf{V}}_{2}}={{\left[ {{U}_{r}},\text{i}{{U}_{z}},\Phi ,{{\Sigma }_{0rr}},\text{i}{{\Sigma }_{0rz}},{{\Delta }_{0r}} \right]}^{\text{T}}}$ and the dimensionless  matrices ${{\mathbf{\overline{M}}}_{1}}$ and ${{\mathbf{\overline{M}}}_{2}}$ are 
\renewcommand{\arraystretch}{2}
\begin{equation} \label{M1n}
{{\mathbf{\overline{M}}}_{1}}=\left[ \begin{matrix}
\dfrac{{{c}_{69}}}{{{c}_{66}}}\dfrac{1}{{{\xi }}} & \dfrac{{{\mu }_{10}}}{{{c}_{66}}}  \\
-{{{\hat{\varpi }}}^{2}}+\dfrac{1}{{{{{\xi }}}^{2}}}\dfrac{{{q}_{7}}}{{{\mu }_{10}}}+\dfrac{{{c}_{77}}}{{{\mu }_{10}}}{{{\hat{\chi }}}^{2}} & -\left( \dfrac{{{c}_{69}}}{{{c}_{66}}}+1 \right)\dfrac{1}{{{\xi }}}
\end{matrix} \right],
\end{equation}
\begin{equation} \label{M2n}
{{\mathbf{\overline{M}}}_{2}}=\left[ \begin{matrix}
-\dfrac{1}{{{\xi }}} & -\hat{\chi } & 0 & 0 & 0 & 0  \\
{{\psi }_{1}}\hat{\chi } & 0 & {{\psi }_{2}}\hat{\chi } & 0 & \dfrac{{{\mu }_{10}}}{{{c}_{55}}} & 0  \\
\dfrac{{{\psi }_{3}}}{{{\xi }}} & {{\psi }_{4}}\hat{\chi } & 0 & 0 & 0 & -\dfrac{{{\varepsilon }_{a0}}}{{{\varepsilon }_{11}}}  \\
-{{{\hat{\varpi }}}^{2}}+\dfrac{1}{{{{{\xi }}}^{2}}}\dfrac{{{q}_{3}}}{{{\mu }_{10}}}+\dfrac{{{q}_{9}}}{{{\mu }_{10}}}{{{\hat{\chi }}}^{2}} & \dfrac{{\hat{\chi }}}{{{\xi }}}\dfrac{{{q}_{4}}}{{{\mu }_{10}}} & {{\psi }_{5}}{{{\hat{\chi }}}^{2}} & 0 & -{{\psi }_{1}}\hat{\chi } & -\dfrac{{{\psi }_{3}}}{{{\xi }}}  \\
\dfrac{{\hat{\chi }}}{{{\xi }}}\dfrac{{{q}_{5}}}{{{\mu }_{10}}} & -{{{\hat{\varpi }}}^{2}}+\dfrac{{{q}_{6}}}{{{\mu }_{10}}}{{{\hat{\chi }}}^{2}} & 0 & {\hat{\chi }} & -\dfrac{1}{{{\xi }}} & -{{\psi }_{4}}\hat{\chi }  \\
{{\psi }_{5}}{{{\hat{\chi }}}^{2}} & 0 & -\dfrac{{{q}_{12}}}{{{\varepsilon }_{a0}}}{{{\hat{\chi }}}^{2}} & 0 & -{{\psi }_{2}}\hat{\chi } & -\dfrac{1}{{{\xi }}}
\end{matrix} \right].
\end{equation}
\renewcommand{\arraystretch}{1}
Here the following dimensionless quantities were introduced,
\begin{equation} \label{dimensionless1}
{{\psi }_{1}}=\frac{{{c}_{58}}}{{{c}_{55}}},\qquad {{\psi }_{2}}=\frac{{{e}_{35}}}{{{c}_{55}}}\sqrt{\frac{{{\mu }_{10}}}{{{\varepsilon }_{a0}}}},\qquad {{\psi }_{3}}={{q}_{1}}\sqrt{\frac{{{\varepsilon }_{a0}}}{{{\mu }_{10}}}},\qquad {{\psi }_{4}}={{q}_{2}}\sqrt{\frac{{{\varepsilon }_{a0}}}{{{\mu }_{10}}}},\qquad {{\psi }_{5}}=\frac{{{q}_{10}}}{\sqrt{{{\mu }_{10}}{{\varepsilon }_{a0}}}},
\end{equation}
as well as 
\begin{equation} \label{dimensionless2}
\hat{\varpi }=\omega a/\sqrt{{{\mu }_{10}}/\rho}=\varpi {{\lambda }_{a}}/({{\eta }}-1),\qquad \hat{\chi }=ka=\chi {{\lambda }_{a}}/({{\eta }}-1),
\end{equation}
where $\varpi =\omega H/\sqrt{{{\mu }_{10}}/\rho}$ and $\chi =kH$ are the dimensionless circular frequency and axial wave number, respectively. Thus, the \emph{dimensionless phase velocity} is defined as ${{v}_{p}}=\varpi /\chi =c/\sqrt{{{\mu }_{10}}/\rho }$, where $c=\omega /k$ is the actual phase velocity.

Equations~\eqref{dimen-state-equation}$_1$ and \eqref{dimen-state-equation}$_2$ are differential systems with variable coefficients, which are intractable analytically and numerically stiff.
To circumvent this difficulty, we employ the approximate laminate model, and divide the tube into $n$ equal thin sublayers with thickness $h/n$ being sufficiently small  that the ${{\mathbf{\overline{M}}}_{k}}$ within each sublayer may be  approximately constant. 
To be specific, the material parameters and the dimensionless deformed radial coordinate itself take the values at the mid-plane of each sublayer.

Accordingly, the formal solutions in the $j$th sublayer can be written as \citep{wu2017guided}
\begin{align} \label{jthlayer}
& {{\mathbf{V}}_{k}}({\xi })=\exp \left[({\xi }-{{{{\xi }}}_{j0}}){{{\mathbf{\overline{M}}}}_{kj}}({{{{\xi }}}_{jm}})\right]{{\mathbf{V}}_{k}}({{{{\xi }}}_{j0}}), \notag \\ 
& \left( {{{{\xi }}}_{j0}}\le {\xi }\le {{{{\xi }}}_{j1}}; \quad k=1,2; \quad j=1,2,\cdots n \right),
\end{align}
where ${{{\mathbf{\overline{M}}}}_{kj}}({{\xi }_{jm}})$ are the approximated constant system matrices within the $j$th sublayer by taking $\xi={\xi }_{jm}$. The dimensionless deformed radial coordinates at the inner, outer and middle surfaces of the $j$th sublayer are
\begin{equation}
{{{\xi }}_{j0}}=1+(j-1)\frac{{{{\overline{\eta }}}}-1}{n},\qquad {{{\xi }}_{j1}}=1+j\frac{{{{\overline{\eta }}}}-1}{n},\qquad {{{\xi }}_{jm}}=1+\frac{(2j-1)({{{\overline{\eta }}}}-1)}{2n},
\end{equation}
respectively, where ${\overline{\eta }}=b/a$. From Eq.~\eqref{jthlayer}, the following recurrence formulas between the state vectors at the inner and outer surfaces of the $j$th sublayer are derived,
\begin{equation} \label{recur}
{{\mathbf{V}}_{k}}({{{\xi }}_{j1}})=\exp \left[({{\overline{\eta }}}-1) {{\mathbf{\overline{M}}}_{kj}}/n\right]{{\mathbf{V}}_{k}}({{{\xi }}_{j0}}),\quad k\in \left\{ 1,2 \right\},
\end{equation}
which, combined with the continuity conditions of state variables at each fictitious interface, results in 
\begin{equation} \label{GTMM}
\mathbf{V}_{k}^{1}={{\mathbf{K}}_{k}}\mathbf{V}_{k}^{0},
\quad k\in \left\{ 1,2 \right\},
\end{equation}
where $\mathbf{V}_{k}^{1}$ and $\mathbf{V}_{k}^{0}$ are the incremental state vectors at the outer and inner surfaces of the tube, respectively, and ${{\mathbf{K}}_{k}}=\prod{_{j=n}^{1}\exp [({{{\overline{\eta}}}}-1){{{\mathbf{\bar{M}}}}_{kj}}/n]}$ is the global transfer matrix of second-order ($k=1$) or sixth-order ($k=2$) through which the boundary state variables at the inner and outer surfaces are connected.


\section{Dispersion relations for axisymmetric waves}
\label{section5}



\subsection{Incremental boundary conditions}
\label{5.1}


Due to the  electric voltage applied at the flexible electrodes, there are no incremental electric fields outside the  tube for the superimposed motion. Furthermore, the internal and external pressures $P_{\text{inn}}$ and $P_{\text{out}}$ as well as the voltage $V$ are kept fixed during the incremental motion, so that the incremental mechanical and electric boundary conditions {\color{black}  Eqs.~\eqref{s1-incremental-boundary}$_{1,2}$ and \eqref{Pcondition}} in cylindrical coordinates read
\begin{align} \label{ME1}
& {{{\dot{T}}}_{0rr}}={{P}_{\text{inn}}}\frac{\partial {{u}_{r}}}{\partial r},\quad {{{\dot{T}}}_{0r\theta }}={{P}_{\text{inn}}}\frac{1}{r}\left( \frac{\partial {{u}_{r}}}{\partial \theta }-{{u}_{\theta }} \right),\quad{{{\dot{T}}}_{0rz}}={{P}_{\text{inn}}}\frac{\partial {{u}_{r}}}{\partial z},\quad \dot{\phi }=0,\quad \left( r=a \right), \notag\\ 
& {{{\dot{T}}}_{0rr}}={{P}_{\text{out}}}\frac{\partial {{u}_{r}}}{\partial r},\quad {{{\dot{T}}}_{0r\theta }}={{P}_{\text{out}}}\frac{1}{r}\left( \frac{\partial {{u}_{r}}}{\partial \theta }-{{u}_{\theta }} \right),\quad {{{\dot{T}}}_{0rz}}={{P}_{\text{out}}}\frac{\partial {{u}_{r}}}{\partial z},\quad \dot{\phi }=0,\quad \left( r=b \right).
\end{align}
For axisymmetric waves ($\partial /\partial \theta =0$), these equations, combined with the incremental incompressibility constraint {\color{black}  Eq.~\eqref{incincom}}, reduce to
\begin{align} \label{ME2}
& {{{\dot{T}}}_{0rr}}=-{{P}_{\text{inn}}}\left( \frac{{{u}_{r}}}{r}+\frac{\partial {{u}_{z}}}{\partial z} \right),\quad {{{\dot{T}}}_{0r\theta }}=-{{P}_{\text{inn}}}\frac{{{u}_{\theta }}}{r},\quad {{{\dot{T}}}_{0rz}}={{P}_{\text{inn}}}\frac{\partial {{u}_{r}}}{\partial z},\quad \dot{\phi }=0,\quad \left( r=a \right), \notag \\ 
& {{{\dot{T}}}_{0rr}}=-{{P}_{\text{out}}}\left( \frac{{{u}_{r}}}{r}+\frac{\partial {{u}_{z}}}{\partial z} \right),\quad {{{\dot{T}}}_{0r\theta }}=-{{P}_{\text{out}}}\frac{{{u}_{\theta }}}{r},\quad {{{\dot{T}}}_{0rz}}={{P}_{\text{out}}}\frac{\partial {{u}_{r}}}{\partial z},\quad \dot{\phi }=0,\quad \left( r=b \right).
\end{align}
Then, substituting Eq.~\eqref{solutions} into Eq.~\eqref{ME2} and using Eq.~\eqref{dimensionless2}$_2$, we arrive at the dimensionless form of incremental boundary conditions,
\begin{align} \label{IBC}
& \Sigma _{0rr}^{0}=-P_{\text{inn}}^{*}\left( U_{r}^{0}+\text{i}U_{z}^{0}\hat{\chi } \right),
&& \Sigma _{0r\theta }^{0}=-P_{\text{inn}}^{*}U_{\theta }^{0},
&& \text{i} \Sigma _{0rz}^{0}=-P_{\text{inn}}^{*}\hat{\chi }U_{r}^{0},
&& {{\Phi }^{0}}=0, \notag \\ 
& \Sigma _{0rr}^{1}=-P_{\text{out}}^{*}\left( \overline{\eta }^{-1}U_{r}^{1}+\text{i}U_{z}^{1}\hat{\chi } \right),
&& \Sigma _{0r\theta }^{1}=-P_{\text{out}}^{*}\overline{\eta }^{-1}U_{\theta }^{1},
&& \text{i}\Sigma _{0rz}^{1}=-P_{\text{out}}^{*}\hat{\chi }U_{r}^{1},
&& {{\Phi }^{1}}=0.
\end{align}


\subsection{Dispersion relations}
\label{5.2}


By combining the incremental boundary conditions {\color{black}  Eq.~\eqref{IBC}} with Eq. \eqref{GTMM}, we obtain two sets of independent linear algebraic equations,
\begin{equation} \label{algeeq}
{{\mathbf{\hat{K}}}_{1}}{{\left[ U_{\theta }^{0},U_{\theta }^{1} \right]}^{\operatorname{T}}}=0,\qquad {{\mathbf{\hat{K}}}_{2}}{{\left[ U_{r}^{0},\text{i}U_{z}^{0},\Delta _{0r}^{0},U_{r}^{1},\text{i}U_{z}^{1},\Delta _{0r}^{1} \right]}^{\operatorname{T}}}=0,
\end{equation}
where ${{\mathbf{\hat{K}}}_{1}}$ and ${{\mathbf{\hat{K}}}_{2}}$ are the coefficient matrices of second-order and sixth-order, respectively, with nonzero components 
\begin{equation} 
(\hat K_1)_{k1} = (K_1)_{k1} - P_\text{inn}^* (K_1)_{k2},\quad 
(\hat K_1)_{12} = -1, \quad 
(\hat K_1)_{22} = P_\text{out}^* \overline{\eta }^{-1},\quad (k=1,2),
\end{equation}
and
\begin{align}
&(\hat K_2)_{k1} = (K_2)_{k1} - P_\text{inn}^* \left[ (K_2)_{k4} + \hat \chi (K_2)_{k5} \right],
 && 
(\hat K_2)_{k2} = (K_2)_{k2} - P_\text{inn}^* \hat \chi (K_2)_{k4}, \notag\\ 
& (\hat K_2)_{k3} = (K_2)_{k6},
 \qquad (k=1, \ldots, 6), 
 && 
 (\hat K_2)_{14} = (\hat K_2)_{25} = (\hat K_2)_{66} = -1,\notag \\ 
&
(\hat K_2)_{44} = P_\text{out}^* \overline{\eta}^{-1},
&& 
(\hat K_2)_{45} = (\hat K_2)_{54} = \hat{\chi }P_\text{out}^*.
\end{align}
in which $(K_k)_{ij}$ are the elements of the global transfer matrix $\mathbf{K}_k$.

For non-trivial solutions of Eq.~\eqref{algeeq}, the determinants of the coefficient matrices must vanish, i.e.,
\begin{equation} \label{dispersionR}
\det\left(\mathbf{\hat K}_1 \right) = 0,
\qquad 
\det\left(\mathbf{\hat K}_2 \right) = 0,
\end{equation}
which are the \emph{dispersion relations} of the two independent classes of axisymmetric waves (T and L waves) propagating in the deformed functionally graded SEA tube under radially inhomogeneous biasing fields.

When there is no external pressure ${{P}_{\text{out}}}=0$, Eqs.~\eqref{algeeq} and \eqref{dispersionR} simplify to
\begin{equation} \label{GTM}
\left[ (K_1)_{21} - P_\text{inn}^* (K_1)_{22} \right]U_{\theta }^{0}=0,\qquad \mathbf{\hat{K}}_{2}^{\text{inn}}{{\left[ U_{r}^{0},\text{i}U_{z}^{0},\Delta _{0r}^{0} \right]}^{\operatorname{T}}}=0,
\end{equation}
and
\begin{equation} \label{breath}
(K_1)_{21} - P_\text{inn}^* (K_1)_{22} = 0,\qquad 
\det\left( \mathbf{\hat{K}}_{2}^{\text{inn}} \right)=0,
\end{equation}
respectively, where $\mathbf{\hat{K}}_{2}^{\text{inn}}$ is the third-order matrix with components
\begin{align}
& (\hat K_2^\text{inn})_{j1}= (K_2)_{k1} - P_\text{inn}^*\left[ (K_2)_{k4} + \hat\chi (K_2)_{k5} \right],\qquad 
(\hat K_2^{\text{inn}})_{j2} = (K_2)_{k2} - \hat{\chi }P_\text{inn}^* (K_2)_{k4}, \notag\\ 
& (\hat K_2^{\text{inn}})_{j3} = (K_2)_{k6}\qquad (j=k-2;\text{ } k=3,4,5).
\end{align}
Analogous dispersion relations apply when the  internal pressure ${{P}_{\text{inn}}}$ is vanishing.

For the incompressible, functionally graded,  Mooney-Rivlin ideal dielectric model {\color{black}  Eq.~\eqref{MR1}}, the nonzero components of the instantaneous electro-elastic moduli tensors are evaluated from Appendix B in \cite{wu2017guided} as
\begin{align} \label{instan}
& {{\mathcal{A}}_{01111}}=\lambda _{\theta }^{-2}\lambda _{z}^{-2}\left[ {{\mu }_{1}}-{{\mu }_{2}}\left( \lambda _{\theta }^{2}+\lambda _{z}^{2} \right) \right]+{{\varepsilon }^{-1}}D_{r}^{2},\qquad {{\mathcal{A}}_{01122}}=-2{{\mu }_{2}}\lambda _{z}^{-2}, \notag\\ 
& {{\mathcal{A}}_{01133}}=-2{{\mu }_{2}}\lambda _{\theta }^{-2},\qquad {{\mathcal{A}}_{02222}}=\lambda _{\theta }^{2}\left[ {{\mu }_{1}}-{{\mu }_{2}}\left( \lambda _{\theta }^{-2}\lambda _{z}^{-2}+\lambda _{z}^{2} \right) \right],\qquad {{\mathcal{A}}_{01221}}={{\mu }_{2}}\lambda _{z}^{-2}, \notag\\ 
& {{\mathcal{A}}_{02233}}=-2{{\mu }_{2}}\lambda _{\theta }^{2}\lambda _{z}^{2},\qquad {{\mathcal{A}}_{03333}}=\lambda _{z}^{2}\left[ {{\mu }_{1}}-{{\mu }_{2}}\left( \lambda _{\theta }^{-2}\lambda _{z}^{-2}+\lambda _{\theta }^{2} \right) \right], \qquad {{\mathcal{A}}_{01331}}={{\mu }_{2}}\lambda _{\theta }^{-2}, \notag\\ 
& {{\mathcal{A}}_{01212}}=\lambda _{\theta }^{-2}\left( {{\mu }_{1}}\lambda _{z}^{-2}-{{\mu }_{2}} \right)+{{\varepsilon }^{-1}}D_{r}^{2}, \qquad {{\mathcal{A}}_{02121}}=\lambda _{\theta }^{2}\left( {{\mu }_{1}}-{{\mu }_{2}}\lambda _{z}^{2} \right), \notag\\ 
& {{\mathcal{A}}_{01313}}=\lambda _{z}^{-2}\left( {{\mu }_{1}}\lambda _{\theta }^{-2}-{{\mu }_{2}} \right)+{{\varepsilon }^{-1}}D_{r}^{2}, \qquad {{\mathcal{A}}_{02323}}={{\mu }_{1}}\lambda _{\theta }^{2}-{{\mu }_{2}}\lambda _{z}^{-2}, \notag\\ 
& {{\mathcal{A}}_{02332}}={{\mu }_{2}}\lambda _{\theta }^{2}\lambda _{z}^{2},\qquad {{\mathcal{A}}_{03131}}=\lambda _{z}^{2}\left( {{\mu }_{1}}-{{\mu }_{2}}\lambda _{\theta }^{2} \right),\qquad {{\mathcal{A}}_{03232}}={{\mu }_{1}}\lambda _{z}^{2}-{{\mu }_{2}}\lambda _{\theta }^{-2}, \notag\\ 
& {{\mathcal{M}}_{0111}}=2{{\varepsilon }^{-1}}{{D}_{r}},\qquad {{\mathcal{M}}_{0122}}={{\mathcal{M}}_{0133}}={{\varepsilon }^{-1}}{{D}_{r}},\qquad {{\mathcal{R}}_{011}}={{\mathcal{R}}_{022}}={{\mathcal{R}}_{033}}={{\varepsilon }^{-1}},
\end{align}
which determine the effective material parameters appearing in Eqs.~\eqref{M1n} and \eqref{M2n}. 
Note that for functionally graded SEA tubes, the material parameters ${\mu }_{1}$, ${\mu }_{2}$ and $\varepsilon$ in Eq.~\eqref{instan} depend on the radial coordinate $R$ and satisfy the affine variation {\color{black}  Eq.~\eqref{gradientPara}}.


\section{Numerical results for wave propagation analysis} \label{Sec6}


We now investigate the effects of material gradient parameters and electro-mechanical biasing fields on the characteristics of small-amplitude axisymmetric wave propagation  in a functionally graded SEA tube. 

{\color{black}The ratio of ${{\mu }_{20}}$ to ${{\mu }_{10}}$ is again  ${{\mu }_{20}}/{{\mu }_{10}}=-0.104$, as in Sec.~\ref{section3.3}}. In addition to the elastic moduli gradients ${{\beta }_{1}}, {{\beta }_{2}}$ and the permittivity gradient $\beta_3$, the dynamic behavior analysis now involves the \emph{functionally graded mass density}, which we take as an \emph{affine} variation as well, in the form $\rho(R)=\rho_0 \left( 1+{{\beta }_{4}}R/A \right)$ with $\beta_4$ being the density gradient. 
For wall thickness such that $\eta=B/A=2$, we must have the restriction $\beta_4>-0.5$ for $\rho(R)$ to be positive. 

We set ${{P}_{\text{inn}}^{*}} ={{P}_{\text{out}}^{*}}=0$ in the numerical calculations, because the effect of the pressure difference on axisymmetric wave propagation has already been discussed in detail by \citet{wu2017guided}, albeit  in a purely elastic functionally graded tube. 

\begin{figure}[h!]
	\centering	
	\includegraphics[width=0.95\textwidth]{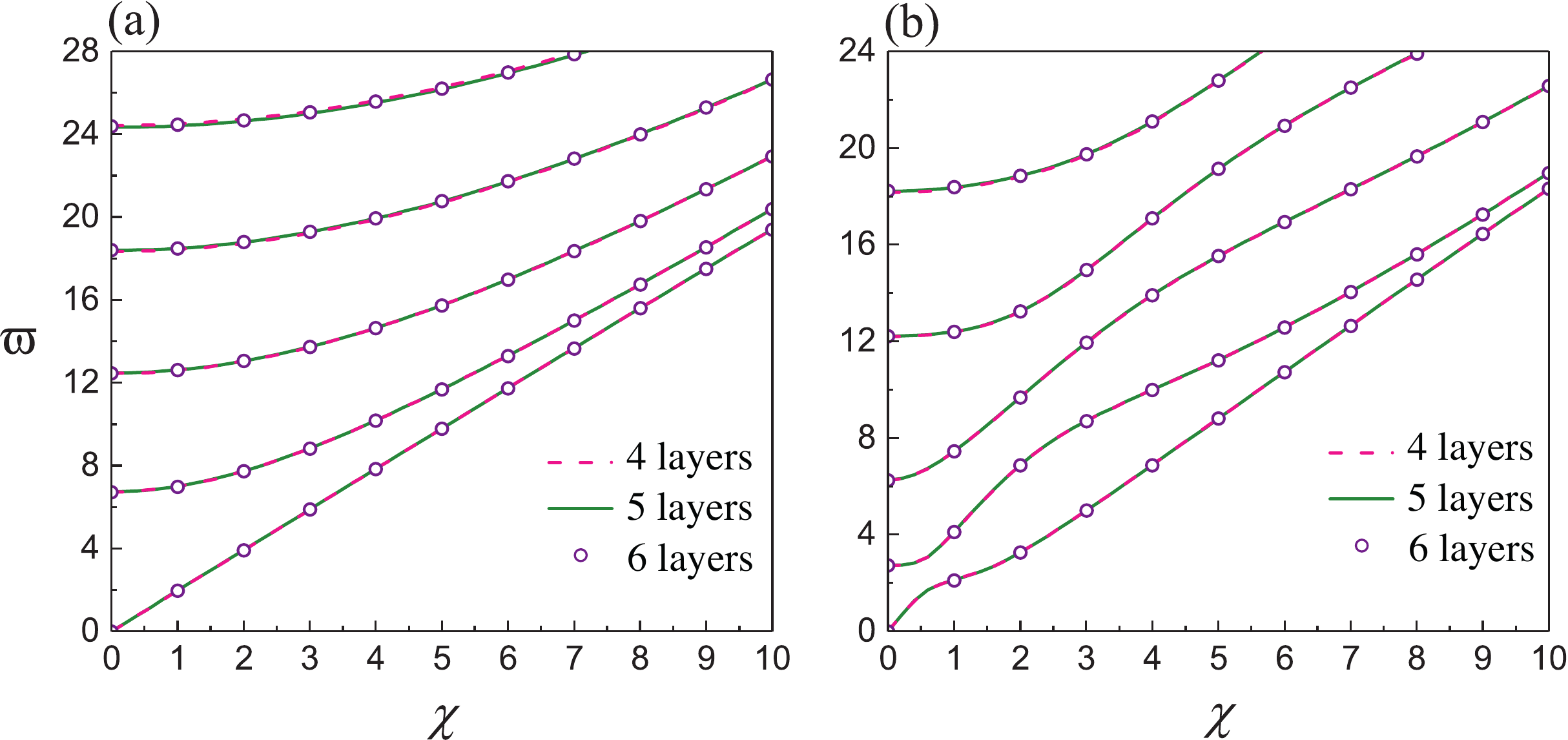}
	\caption{Convergence analysis of frequency spectra of the first five wave modes obtained by the SSM for a functionally graded SEA tube with $\beta_1=\beta_2=5.0$, $\beta_3=\beta_4=1.0$, $\lambda_z=1.0$ and $V^*=0.2$: (a) T waves; (b) L waves. Here $n = 4,5,6$ is the number of the discretized thin layers.}
	\label{Fig9}
\end{figure}


\subsection{Validation of state-space method}


We first examine the effectiveness of the SSM in terms of  accuracy and convergence. 
For a \emph{homogeneous} pre-stretched tube without the action of voltage and pressure difference, the accuracy of SSM to predict the frequency and phase velocity spectra was validated by \citet{wu2018propagation}, by making a comparison with the exact solutions based on the displacement method \citep{su2016propagation}, and thus that analysis is omitted here.

Fig.~\ref{Fig9} displays the frequency spectra of the T and L waves calculated by the SSM for a functionally graded SEA tube subject to the voltage $V^*=0.2$, and for different numbers of  discretized thin layers. {\color{black}We see from Fig.~\ref{Fig9} that the frequency spectra of both T and L waves for the thin-layer number $n=6$ coincide with those corresponding to $n=5$. We also calculate the frequency spectra for the thin-layer number $n=20$, which are almost the same as those for $n=5$ in Fig.~\ref{Fig9} and omitted here for brevity. Thus,} an excellent convergence of the SSM to predict the frequencies is already achieved when the number of thin layers  is $n=5$. Thus, we take 5 layers from now on, which is assumed to have high accuracy. 
In addition, Fig.~\ref{Fig9} shows that the first branch (i.e., the fundamental mode) of the T waves is \emph{almost} non-dispersive despite the presence of material gradient and voltage. The fundamental mode of the L waves, however, is obviously dispersive and it is almost a straight line only at a large wave number. In fact, the fundamental modes of the T and L waves asymptotically tend to the surface shear wave and the Rayleigh-type surface wave, respectively, at a large wave number.


\subsection{Torsional waves}\label{6.2}


\begin{figure}[h!]
	\centering	
	\includegraphics[width=0.7\textwidth]{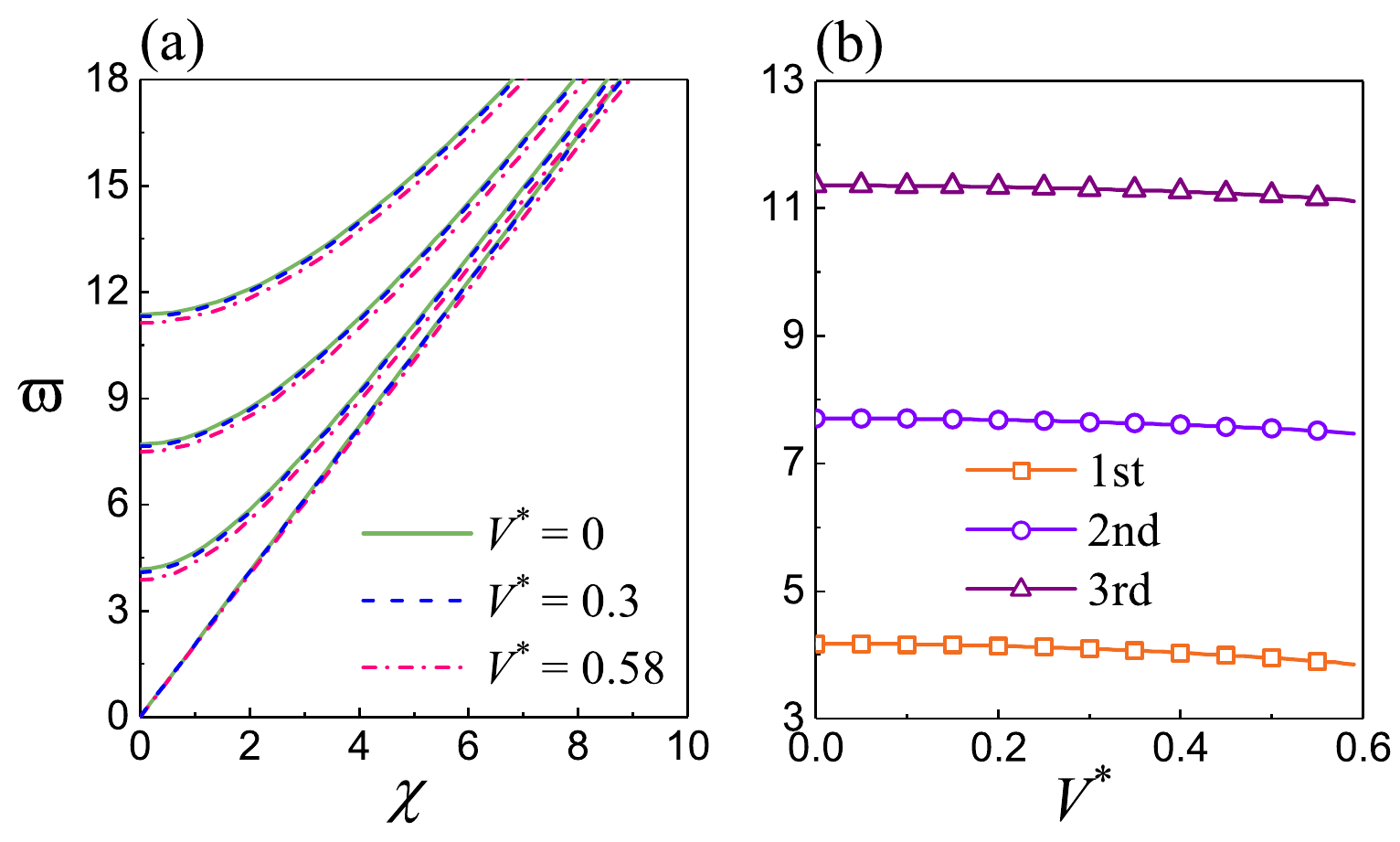}
	\caption{(a) Frequency spectra of the first four modes of the T waves for different values of $V^*$, and (b) the first three nonzero cut-off frequencies at $\chi=0$ versus $V^*$ in a functionally graded SEA tube with $\beta_i=2.0 \text{ } (i=1, \ldots,4)$ and $\lambda_z=2.0$.}
	\label{Fig10}
\end{figure}

Fig.~\ref{Fig10} displays the effect of the applied voltage on the frequency spectra and the  cut-off frequencies at $\chi=0$ of the T waves in a pre-stretched,  functionally graded SEA tube with $\beta_i=2.0 \text{ } (i=1, \ldots, 4)$ and $\lambda_z=2.0$. 
To avoid the collapse of the tube, the applied voltage cannot surpass the critical value $V_c^*\simeq 0.59$ found when $\beta_i=2.0$ (see Fig.~\ref{Fig5}(c)). 

According to Fig.~\ref{Fig10}, the frequency spectra, including the cut-off frequencies, are hardly changed  by the application of a voltage in the entire wave number range, even as the voltage approaches the critical value $V_c^*$. 
This is physically understandable, as the applied direction of the electric field due to the voltage is perpendicular to both the particle motion direction and the propagation direction of the T wave modes, so that  we expect the work done by the biasing electric field to be negligible. 
Confirming numerically that \emph{the frequency is almost independent of the voltage} is a feature that could be exploited to design a torsional waveguide which may work robustly, with a consistent working performance.

\begin{figure}[h!]
	\centering	
	\includegraphics[width=0.95\textwidth]{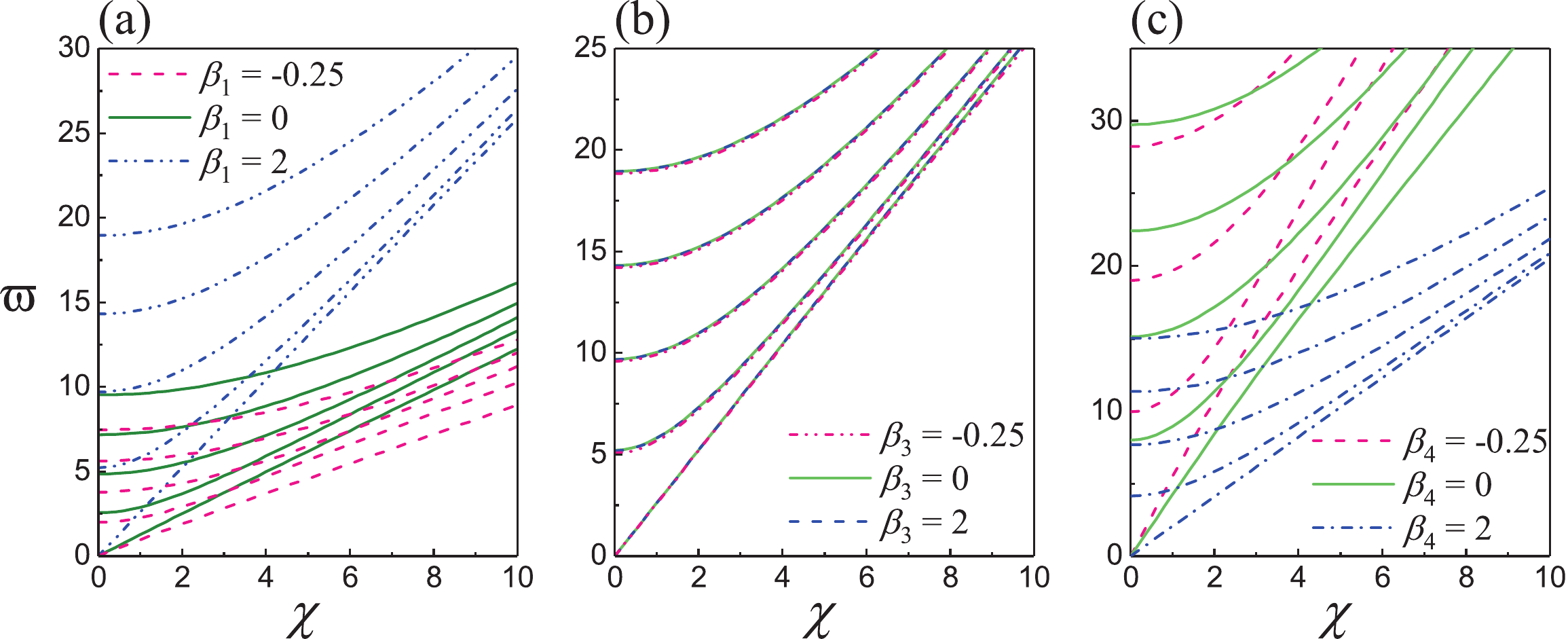}
	\caption{Frequency spectra of the first five modes of the T waves in a functionally graded SEA tube with $\lambda_z=2.0$ and $V^*=0.4$: (a) for various \emph{elastic moduli} gradients $\beta_1=\beta_2$ (with $\beta_3=-0.25$ and $\beta_4=1.0$); (b) for various \emph{permittivity} gradients $\beta_3$ (with $\beta_1=\beta_2=2.0$ and $\beta_4=1.0$); (c) for various \emph{density} gradients $\beta_4$ (with $\beta_1=\beta_2=2.0$ and $\beta_3=0$).}
	\label{Fig11}
\end{figure}

\begin{figure}[htbp]
	\centering	
	\includegraphics[width=0.95\textwidth]{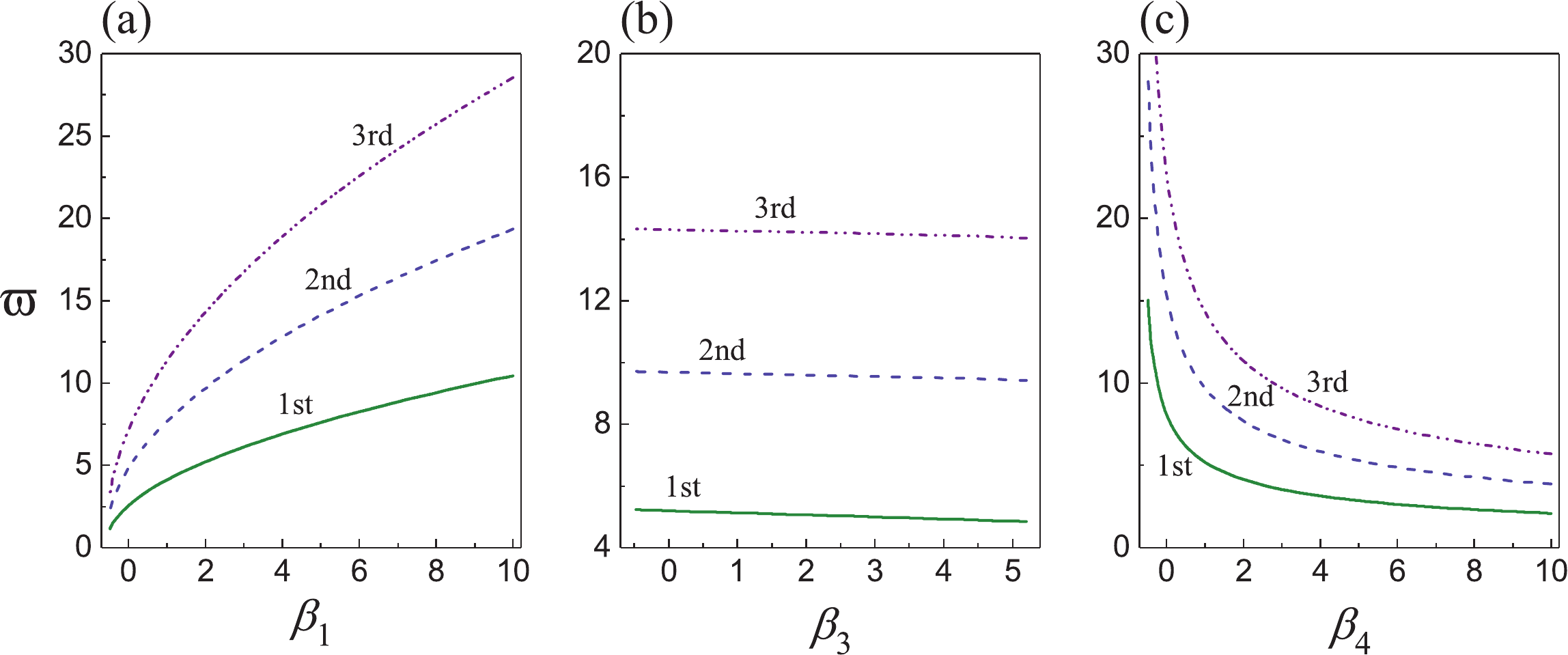}
	\caption{The first three nonzero cut-off frequencies at $\chi=0$ of the T waves versus the gradient parameter in a functionally graded SEA tube with $\lambda_z=2.0$ and $V^*=0.4$: (a) for varying \textit{elastic moduli} gradient $\beta_1=\beta_2$ (with $\beta_3=-0.25$ and $\beta_4=1.0$); (b) for varying \textit{permittivity} gradient $\beta_3$ (with $\beta_1=\beta_2=2.0$ and $\beta_4=1.0$); (c) for varying \textit{density} gradient $\beta_4$ (with $\beta_1=\beta_2=2.0$ and $\beta_3=0$).}
	\label{Fig12}
\end{figure}

Fig.~\ref{Fig11} depicts the frequency spectra of the first five modes of the T waves in a pre-stretched functionally graded SEA tube with $\lambda_z=2.0$ and $V^*=0.4$, for different material gradient values. To further demonstrate the dependence of wave characteristics on the material gradient parameters, Figs.~\ref{Fig12}(a)-(c) show the variations of the first three cut-off frequencies at $\chi=0$ with the elastic moduli gradient $\beta_1=\beta_2$, permittivity gradient $\beta_3$, and density gradient $\beta_4$, respectively. 

We observe from Figs.~\ref{Fig11}(a) and \ref{Fig12}(a) that the cut-off frequencies and the curve slopes of all T wave modes (i.e., the group velocities) have a significant and monotonous rise with  increasing elastic moduli gradient in the entire wave number range. Furthermore, the gap between two neighboring branches becomes larger as $\beta_1$ increases. In fact, increasing the elastic moduli gradient stiffens the  tube, resulting in a remarkable increase in the frequency. On the contrary, Figs.~\ref{Fig11}(c) and \ref{Fig12}(c) reveal that the  frequencies and the group velocities undergo a sharp drop with an increase in the density gradient $\beta_4$. This is because the larger the density gradient is, the lower the frequency is. 
Finally, it is clear from Figs.~\ref{Fig11}(b) and \ref{Fig12}(b) that the frequency spectra, including the cut-off frequencies, are hardly influenced by the permittivity gradient $\beta_3$; this is essentially analogous to the minor effect that the voltage has on the frequency, as seen  in Fig.~\ref{Fig10}.

\subsection{Longitudinal waves}\label{6.3}


We now turn our attention to the investigation of the L waves. 

\begin{figure}[h!]
	\centering	
	\includegraphics[width=0.95\textwidth]{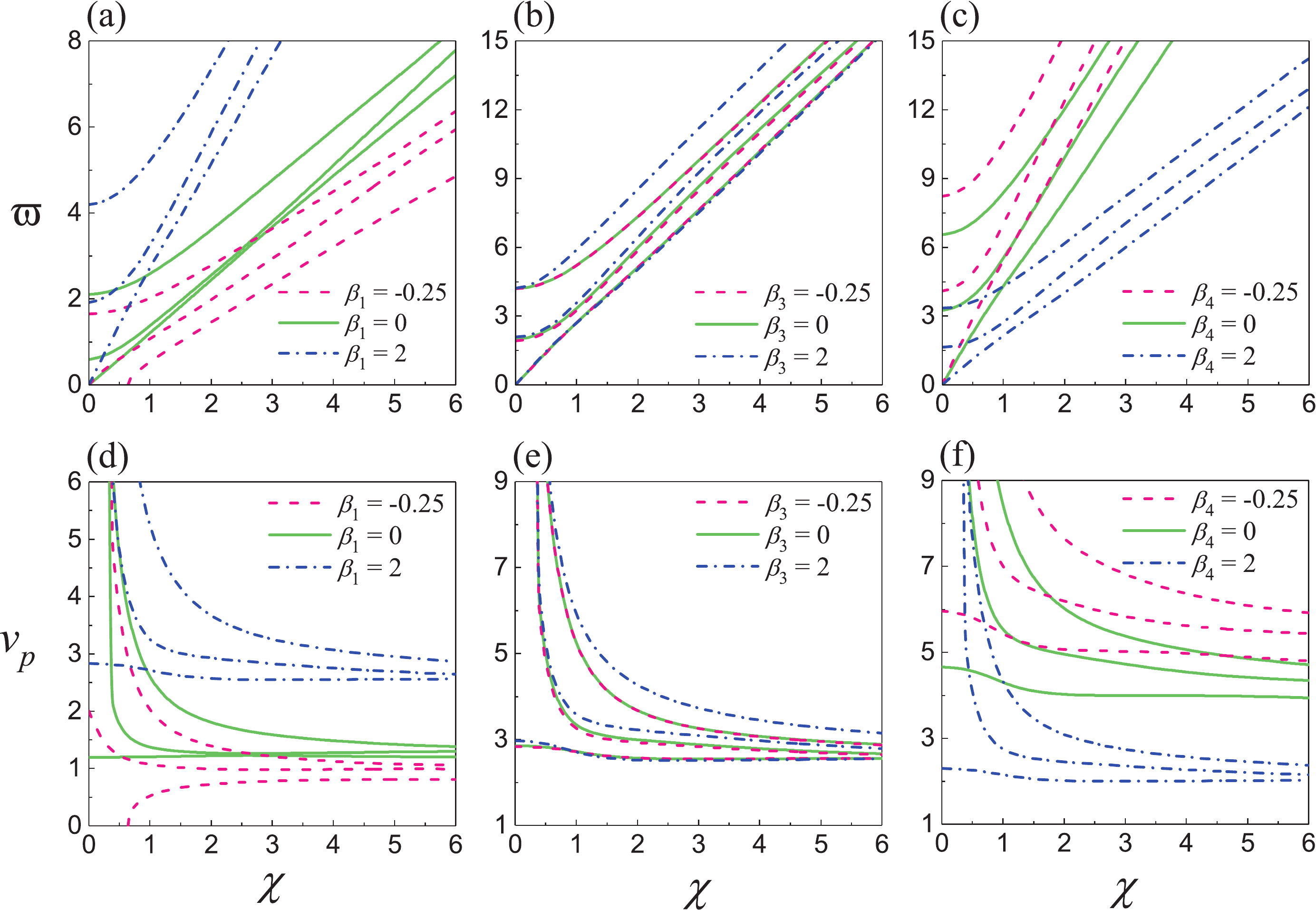}
	\caption{Frequency spectra (a-c) and phase velocity spectra (d-f) of the first three modes of the L waves in a functionally graded SEA tube with $\lambda_z=2.0$ and $V^*=0.4$: (a, d) for various \textit{elastic moduli} gradients $\beta_1=\beta_2$ (with $\beta_3=-0.25$ and $\beta_4=1.0$); (b, e) for various \textit{permittivity} gradients $\beta_3$ (with $\beta_1=\beta_2=2.0$ and $\beta_4=1.0$); (c, f) for various \textit{density} gradients $\beta_4$ (with $\beta_1=\beta_2=2.0$ and $\beta_3=0$).}
	\label{Fig13}
\end{figure}

For fixed biasing fields ($\lambda_z=2.0$ and $V^*=0.4$), Fig.~\ref{Fig13} presents the frequency spectra (Figs.~(a)-(c)) and phase velocity spectra (Figs.~(d)-(f)) of the first three L wave modes for different values of material gradient parameter. 
We see that all the L wave modes are dispersive in all cases (i.e., the phase velocity $v_p$ varies with the wave number $\chi$) except for the fundamental mode when $\beta_1=\beta_2=0$, which is almost non-dispersive with a constant phase velocity (see Fig.~\ref{Fig13}(d)).

It is clear from Fig.~\ref{Fig13} that the material gradients affect significantly the frequency spectra and phase velocity spectra. 
Hence, the frequencies and phase velocities of all L wave modes are lifted up with an increase in the \textit{elastic moduli gradients} $\beta_1=\beta_2$ as a result of the enhanced stiffening effect.
Conversely, the frequencies and phase velocities exhibit a monotonically decreasing trend in the whole wave number range when the \emph{density gradient} $\beta_4$  increases,  due to the increasing mass effect. 
Additionally, increasing $\beta_1=\beta_2$ or decreasing $\beta_4$ also raises remarkably the curve slope (hence, the group velocity) of all L wave modes. 
These phenomena are qualitatively the same as those observed in Figs.~\ref{Fig11}(a) and \ref{Fig11}(c) for the T waves. 
Similar to the independence of the T wave frequency spectra with respect to the {permittivity gradient} $\beta_3$ shown in Fig.~\ref{Fig11}(b), the frequency and phase velocity of the \textit{fundamental} L wave mode remain almost unchanged when varying $\beta_3$. 
For the higher-order L wave modes, the cut-off frequencies are hardly affected by $\beta_3$, but the frequencies and phase velocities gradually increase with the increasing $\beta_3$, especially at a large wave number (see Figs.~\ref{Fig13}(b) and \ref{Fig13}(e)). Thus, \emph{tailoring the material gradient behavior can be used to adjust  the characteristics of elastic wave propagation in functionally graded SEA tubes}.

Fig.~\ref{Fig13} also illustrates that the phase velocity of the higher-order L wave modes originates from infinity with a finite cut-off frequency depending on the gradient parameters as described above. 
Nevertheless, the phase velocity of the fundamental mode in all cases emanates from a finite value which is also associated with the material gradients. 
Interestingly, a cut-off wave number $\chi_{\text{co}}^{\text{anti}} \simeq0.64$ exists in the fundamental mode of the frequency/phase velocity spectra when $\beta_1=\beta_2=-0.25$, that is, the first branch emerges from $\chi_{\text{co}}^{\text{anti}}$. 
Through numerical calculations, the value of $\chi_{\text{co}}^{\text{anti}}$ increases monotonically with a decrease in the elastic moduli gradient to $-0.5$. 
Thus, a critical wavelength is defined by $\chi_{\text{co}}^{\text{anti}}$, below which there is no stable propagation of the longer L waves. This phenomenon is reminiscent of the first antisymmetric Rayleigh-Lamb wave mode in SEA plates subject to biasing fields \citep{shmuel2012rayleigh}.

\begin{figure}[h!]
	\centering	
	\includegraphics[width=1.0\textwidth]{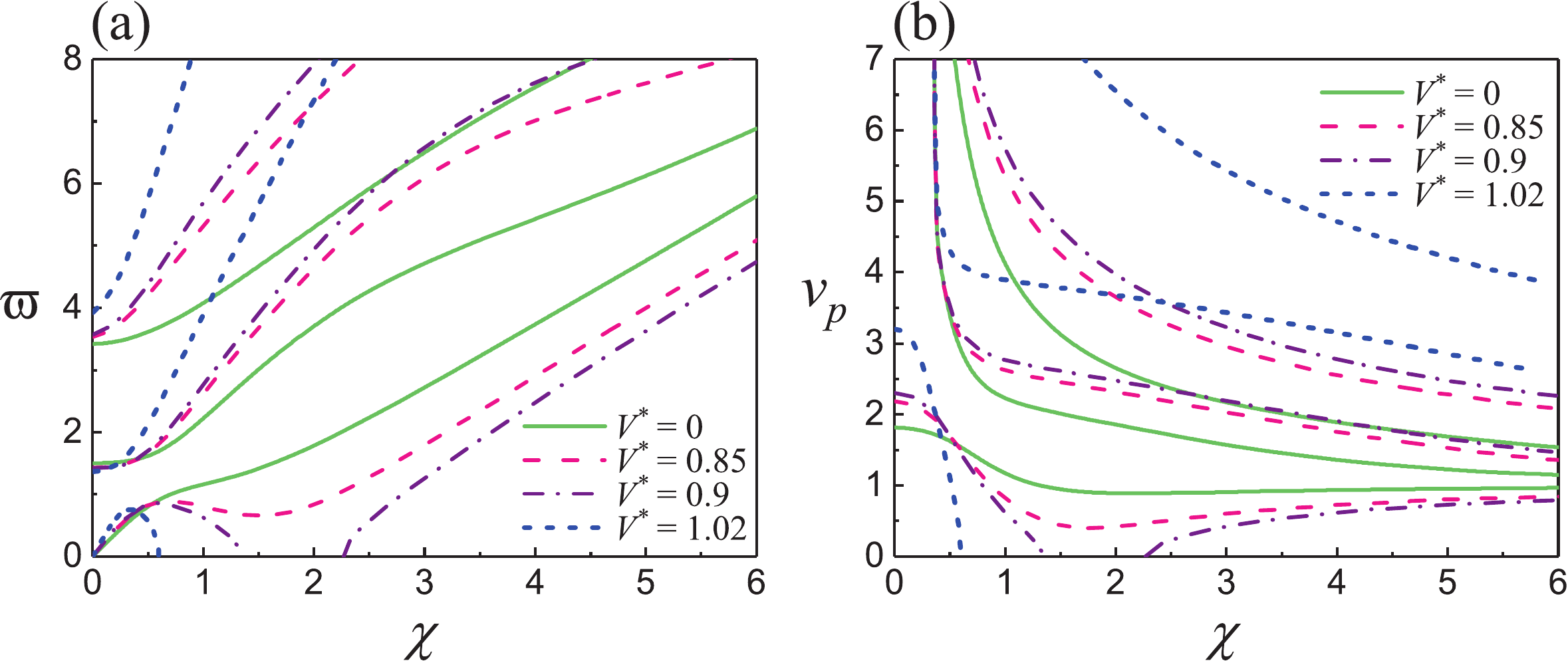}
	\caption{Frequency spectra (a) and phase velocity spectra (b) of the first three modes of the L waves in a functionally graded SEA tube with $\beta_i=2.0 \text{ } (i=1, \ldots, 4)$ and $\lambda_z=1.0$, for different values of $V^*$.}
	\label{Fig14}
\end{figure}

Fig.~\ref{Fig14} demonstrates the effect of the applied voltage on the frequency spectra and phase velocity spectra of the first three L wave modes in a functionally graded SEA tube with $\beta_i=2.0 \text{ } (i=1, \ldots, 4)$ and $\lambda_z=1.0$. 
All the applied voltages are in the allowable range (i.e., below the critical value for collapse). 
We observe that the frequencies and phase velocities for the higher-order modes have a monotonically increasing behavior with an increase in $V^*$, except for the first nonzero cut-off frequency which remains almost unchanged. 

\begin{figure}[h!]
	\centering	
	\includegraphics[width=0.95\textwidth]{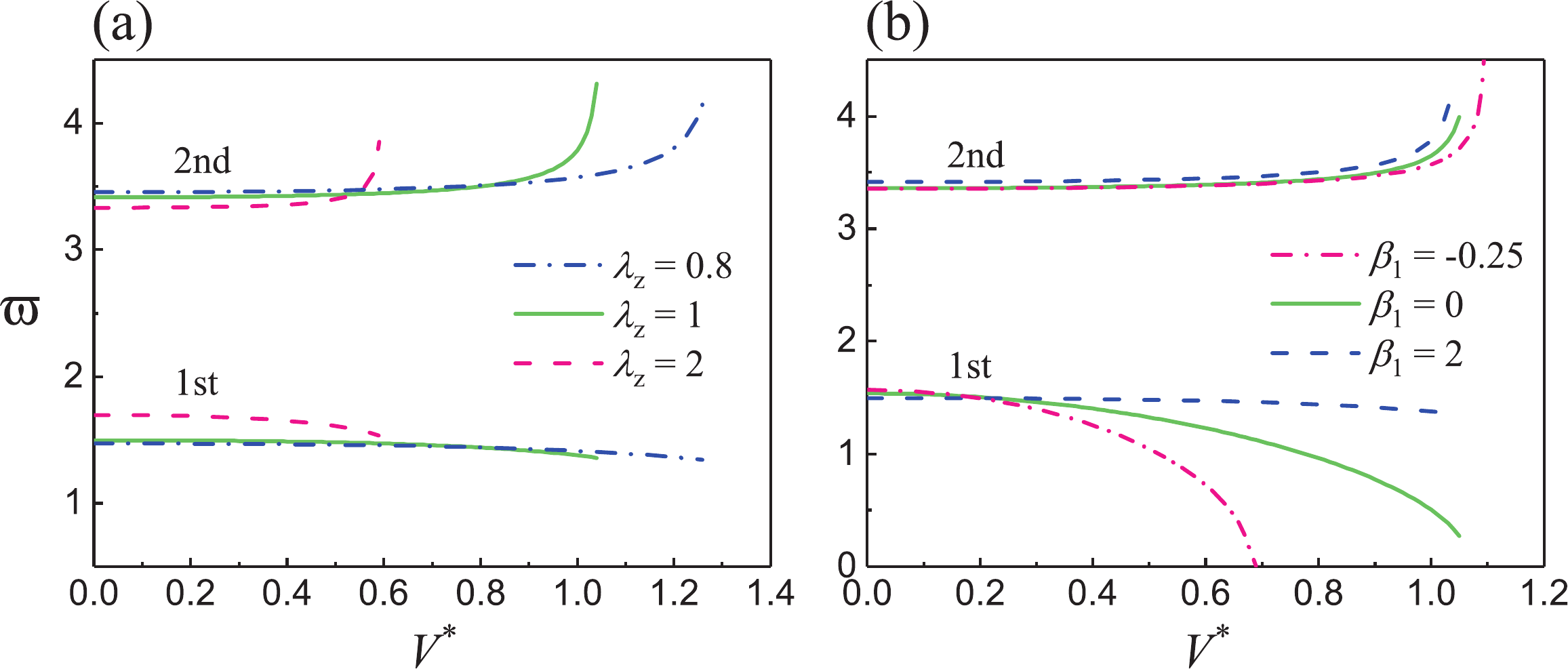}
	\caption{The first two nonzero cut-off frequencies at $\chi=0$ of the L waves versus $V^*$: (a) for different axial pre-stretches $\lambda_z$ (with $\beta_i=2.0,\, i=1, \ldots, 4$); (b) for different material gradients $\beta=\beta_i \text{ } (i=1, \ldots, 4)$ (with $\lambda_z=1.0$).}
	\label{Fig15}
\end{figure}

Fig.~\ref{Fig15} clearly illustrates the dependence of the first two nonzero cut-off frequencies on the applied voltage for different axial pre-stretches and material gradient parameters determining various critical voltages (see Fig.~\ref{Fig5}). 
We observe from Fig.~\ref{Fig15}(a) that for the three axial pre-stretches $\lambda_z=0.8, 1.0, 2.0$, the first cut-off frequency has a slight decrease with the voltage while the second cut-off frequency increases gradually with the voltage. 
Fig.~\ref{Fig15}(b) reveals that varying the material gradient does not alter the variation trend of the second cut-off frequency with the voltage. The first cut-off frequency, however, declines continually with the voltage when the material gradient drops down.

Furthermore, Fig.~\ref{Fig14} indicates that the influence of the voltage on the frequency/phase velocity spectra of the \emph{fundamental mode} is more complex than for the higher-order modes. 
For $V^*=0$, the frequency of the fundamental mode increases continuously with the wave number while its phase velocity declines monotonically. When the voltage increases to a high value such as $V^*=0.85$, the frequency and phase velocity exhibit a non-monotonic variation with the wave number. Specifically, the negative slope of the frequency curve emerges within a certain wave number range and then becomes positive once again. This is a peculiar phenomenon indicating that in response to the same excitation frequency, there exist more than one wave with various wavelengths and velocities propagating in the fundamental branch. In addition, the phase velocity of the fundamental mode first decreases to a minimum value and subsequently increases gradually with the wave number, asymptotically tending to the modified Rayleigh surface wave velocity \citep{broderick2020electro}. This phenomenon results from the complex wave interaction with the geometric boundaries in terms of the thickness and mean radius of the deformed  tube \citep{shmuel2013axisymmetric, wu2017guided}. 
Further increase in $V^*$ leads to a finite cut-off wave number of the fundamental frequency branch with negative slope, and a new branch emerges at another finite wave number. This phenomenon is discussed in detail below.

\begin{figure}[htbp]
	\centering	
	\includegraphics[width=1.0\textwidth]{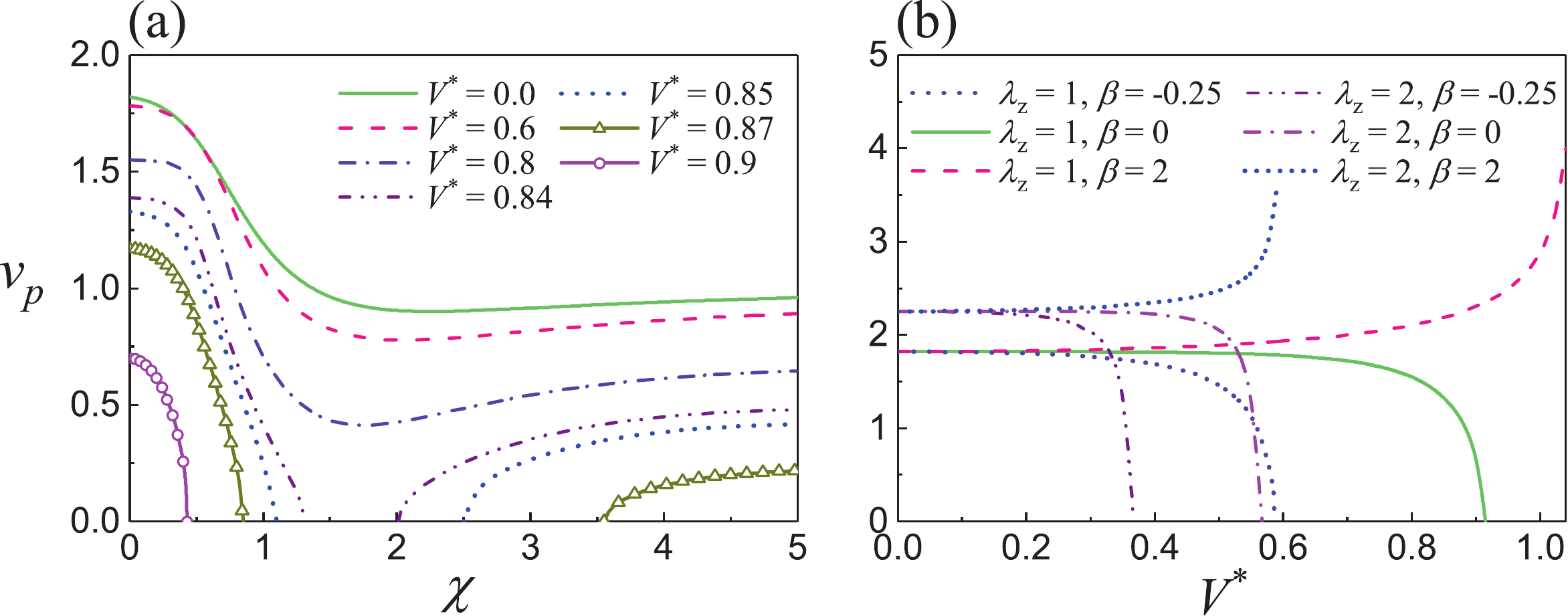}
	\caption{(a) Phase velocity spectra of the lowest L wave mode in a \emph{homogeneous} ($\beta_i=0$, $i=1, \ldots, 4$) SEA tube with $\lambda_z=1.0$ for different values of $V^*$. (b) Variations of the lowest L wave phase velocity at $\chi=0$ with $V^*$ for different combinations of the axial pre-stretch $\lambda_z$ and material gradient $\beta=\beta_i \text{ } (i=1, \ldots, 4)$.}
	\label{Fig16}
\end{figure}

For a homogeneous ($\beta_i=0$) SEA tube with $\lambda_z=1.0$, Fig.~\ref{Fig16}(a)  presents the evolution of the phase velocity spectra of the fundamental branch for different values of voltage. 
The variation trend of $v_p$ with $\chi$ in the voltage range below the critical value $V_{c}^*\simeq0.94$ is qualitatively similar to that shown in Fig.~\ref{Fig14}(b) for the functionally graded tube with $\beta_i=2.0$. However, comparing Fig.~\ref{Fig16}(a) with Fig.~\ref{Fig14}(b), we find that the phase velocity has a monotonous decrease with an increase in voltage $V^*$ over the entire wave number range when $\beta_i=0$ (no gradient), while for the functionally graded tube with $\beta_i=2.0$, the phase velocity increases gradually with $V^*$ at the small wavenumber (i.e., long wavelength) range less than a certain wave number. 
These phenomena are also observed in Fig.~\ref{Fig16}(b) for the variation of the lowest phase velocity at $\chi=0$ with $V^*$ for different combinations of $\lambda_z$ and $\beta_i$, where the axial pre-stretch does not change the variation trend. Thus, in principle, \emph{in-situ ultrasonic nondestructive evaluation can be utilized to characterize the material gradient behavior and the operating biasing fields state}.

Similar to the functionally graded case in Fig.~\ref{Fig14}(b), Fig.~\ref{Fig16}(a) shows that for a large enough value of voltage (for example, $V^* \geqslant 0.84$) and as the wave number increases, the phase velocity 
first decreases gradually to zero at a smaller finite wave number $k_1^{\text{sym}}$ and subsequently the fundamental branch emerges again at another larger wave number $k_2^{\text{anti}}$. This particular phenomenon also emerges in the study of symmetric/antisymmetric Rayleigh-Lamb waves propagating in a homogeneous SEA plate \citep{shmuel2012rayleigh}. In fact, the circumferential stretch $\lambda_a$ of the tube increases considerably for a high voltage as observed in Fig.~\ref{Fig5}, leading to a remarkable increase in the curvature radius. At this moment, there is no difference between the tube and the flat plate for the L wave fundamental mode, and hence, the L wave propagation behaviors resemble those of the Rayleigh-Lamb wave in a SEA plate. As a result, this behavior provides a possibility to annihilate the L wave propagation in the wavelength range corresponding to the interval $(k_1^{\text{sym}}, k_2^{\text{anti}})$. Furthermore, Fig.~\ref{Fig16}(a) also shows that raising the applied voltage results in a decrease in $k_1^{\text{sym}}$ and a rapid increase in $k_2^{\text{anti}}$, and thus enlarges the range of the annihilated wavelength. When the value of $k_2^{\text{anti}}$ tends to infinity due to the increasing voltage, the zero phase velocity represents the surface instability of a SEA half-space subject to an electric field \citep{dorfmann2010nonlinear, su2018wrinkles}.


\section{Conclusions}\label{sec7}


We presented a theoretical analysis of  finite axisymmetric deformation and  superimposed axisymmetric wave propagation in a functionally graded SEA tube subject to electro-mechanical biasing fields. We derived the explicit expressions governing the static finite deformation and the radially inhomogeneous biasing fields in the tube for the generalized Mooney-Rivlin ideal dielectric model with a radial affine gradient variation. Employing the state-space method in cylindrical coordinates, we obtained analytically the dispersion relations for the small-amplitude T and L waves propagating in the deformed tube. Finally, we conducted detailed calculations to elucidate the dependence of the static nonlinear response as well as the T and L wave propagation behaviors on the biasing fields and material gradient parameters. Our numerical findings demonstrate that (i) tailoring properly the gradation of material properties may improve the actuation performance by a low voltage and alleviate the stress inhomogeneity in SEA actuators; (ii) the axisymmetric wave behaviors in SEA tubes may be readily tuned via adjusting the biasing fields and material gradient properties; and (iii) the material properties and working state of SEA tubes may be characterized by  real-time ultrasonic nondestructive testing.

{\color{black}We only considered the uncoupled T and L axisymmetric guided waves. Other propagation modes include non-axisymmetric waves and circumferential guided waves, and are worthy of further research. 

The SEA tube studied in this work is characterized by the generalized functionally graded Mooney-Rivlin ideal dielectric model, which does not capture the strain-stiffening effect \citep{destrade2017methodical}. Provided the tube is not deformed excessively, the Mooney-Rivlin model provides a general framework for this exploration, because it is equivalent to all isotropic material models in the small-to-moderate regime of deformations \citep{rivlin1951large, destrade2010onset}.
The strain-stiffening effect of other nonlinear material models \citep{ mangan2015gent, destrade2017methodical} on the wave propagation characteristics of functionally graded SEA tubes remains to be explored.}

{\color{black}Finally, we point out that  nonreciprocal transmission of acoustic/elastic waves can be achieved in principle based on  structural asymmetry or material nonlinearity \citep{li2018nonreciprocal, li2019diode, chen2019tunable}. 
Using functionally graded SEA materials to design actively tunable acoustic/elastic diodes is an  interesting topic to address in the future.}


\section*{Acknowledgments}


This work was supported by a Government of Ireland Postdoctoral Fellowship from the Irish Research Council (No. GOIPD/2019/65) and the National Natural Science Foundation of China (Nos. 11872329 and 11621062). Partial supports from the Fundamental Research Funds for the Central Universities (No. 2016XZZX001-05) and the Shenzhen Scientific and Technological Fund for R \& D (No. JCYJ20170816172316775) are also acknowledged. Michel Destrade thanks Zhejiang University for organising research visits.
 

\section*{References}

\bibliographystyle{elsarticle-harv.bst}
\nocite{*}
\bibliography{FG_DE_Cylindrical_Shell-v2.bib}







\end{document}